%% file: E2.tex
\font\eightrm=cmr8 at 8pt
\documentclass{oxarticle}
\usepackage{amsmath, amssymb}
\usepackage{graphics, setspace}
\usepackage{epstopdf}
\usepackage{bm}
\usepackage{tikz}
 \input{Johnmacros27}

\newcounter{orange} 
\newcounter{apple} 
\addtocounter{apple}{1}
\newcounter{grape} 
\addtocounter{grape}{1}
\newcommand{\numberhere}{EPapersNew}
\newcommand{\articlenumber}{e2}

\proofmodefalse
\usepackage{color}

\newcommand{\mathsym}[1]{{}}
\newcommand{\unicode}[1]{{}}

\begin{document}

 \begin{center}

{  \Large  The BRS Cohomology of the  Wess Zumino Chiral Scalar\\ supersymmetric model \\ with exotic pairs and exotic triplets (E2)
 \\[1cm]}  
%

\renewcommand{\thefootnote}{\fnsymbol{footnote}}

{\Large John A. Dixon\footnote{jadixg@gmail.com, john.dixon@ucalgary.ca}\\Physics Dept\\University of Calgary \\[1cm]}  
\end{center}
 
 \Large

 \begin{center}
Abstract
\end{center}

Using the spectral sequence method, this paper advances some of the construction of the BRS cohomology of the Wess Zumino supersymmetric action.  An important missing part was the inclusion of the sources for the variations of the fields.  In this paper, these sources are called pseudofields.  Since the most interesting part of the result  contains unsaturated spinor indices, we include a constant spinor $\f_{\dot \a}$ to saturate those indices.   At dimension zero, this gives rise to a new set of invariants and a closely related new set of possible supersymmetry anomalies in the theory, and we call this an `exotic pair'.   At dimension one, this becomes more complicated, and the theory adds a new ghost charge - 1 term, which we call a change,  and we call this an `exotic triplet'.  For higher dimension and higher spin, it appears that more complications are likely to occur. 

 These exotic pairs and triplets are constrained by some simple equations which arise from the spectral sequence.  The invariants  of the exotic pairs are all dependent on the pseudofields, which means that the field parts of these invariants are not supersymmetric, though the invariants are in the cohomology space of supersymmetry. In this paper we examine the BRS cohomology for  spins $0, \fr{1}{2}$ and low dimensions.

\section{\LARGE Introduction}
\la{onesec}
\subsection{\Large Does SUSY have its own type of Anomalies?}

   \refstepcounter{orange}
{ \theorange}.\;{ Does supersymmetry have its own set of anomalies?} Can supersymmetry exhibit anomalies that relate to  the supersymmetry itself?  
Since the beginning of supersymmetry in 3+1 dimensions, starting  with the model of Wess and Zumino \ci{WZ} in 1974, it has been believed that  SUSY  has nothing more than supersymmetric versions of the known gauge types of  anomalies.    The result of this paper is that the BRS cohomology of the   model of Wess and Zumino shows that it is very likely that SUSY has a vast number of anomalies that are totally unrelated to the known gauge type anomalies. These new anomalies  were  introduced, rather informally, in the recent paper \ci{dixexotic}.  That paper did not anticipate the exotic triplets, which emerge in section \ref{nightmare} only after quite a lot of work with the spectral sequence.

 \refstepcounter{orange}
{ \theorange}.\;{ The result proved in this paper is as follows:} 
supersymmetry has  unusual BRS cohomology, because the cohomology contains  `exotic pairs' and `exotic triplets'. We will denote them by $(\cE, \W)$ and $(\cC, \cE, \W)$ respectively, and refer to them as `exotic sets'. All the exotic sets have unsaturated spinor indices, so they do not appear in the usual action, which is one of the reasons they have remained unnoticed for 50 years. 
 In the present work, these exotic sets are exhibited by 
coupling\footnote{In a paper being prepared it will be shown that these indices can be contracted with \cdss \ multiplets, which appears to generate a new direction for the examination of SUSY theories. There also seems to be  much more structure  with multiple spinor indices.}  them to a constant spinor $\ov \f_{\a}$ .   There are several unusual features about the exotic sets:
\ben
\item
They have new invariants of the cohomology, which we will denote by $\cE$. These have ghost charge zero.  However these invariants are unusual because they are composed of two parts $\cE= \cE_1 + \cE_2$ which are summed together:
\ben
\item
The part $\cE_1$ is linear in the  source pseudofields, accompanied by fields and ghosts
\item
The part $\cE_2$   is made of fields without pseudofields or ghosts.
\een
\item
The part $\cE_1$  which contains pseudofields has a variation which vanishes by the Field equations of motion, arising from  the variation of the pseudofields. 
\item
The part $\cE_2$, which is made of fields, is not invariant under supersymmetry.  Its variation cancels the variation of $\cE_1$, and that variation turns out to also be proportional to the field equations. 
\item
In addition, each exotic pair $(\cE, \W)$ has a new `anomaly' part $\W$ that has ghost charge one.  These are possible anomalies of the theory, but a Feynman diagram calculation is needed to see what the coefficients are.  
\item
In addition, each exotic triplet  $(\cC, \cE, \W)$ has a new `Change' part 
$\cC$ that has ghost charge minus one.  These are possible changes to  the theory .  
\item
The exotic   pairs  also have constraint equations.  These are as follows:
\be
d_2 \cE=0
\ee
\be
d_2^{\dag} \W=0
\ee
\item
The exotic   tripets   also have constraint equations for all three  parts.  These are as follows:
\be
d_2 \cE=0
\ee
\be
d_2^{\dag} \W=0
\ee
\be
d_3  \cC=0
\ee
\be
d_3^{\dag} \cE=0
\ee
\item Each of the two pieces of the pair $(\cE, \W)$  has the same dimension and quantum numbers, except that $\cE$ has ghost charge zero and $\W$ has ghost charge one.  These are derived in section \ref{dimzerosec} below.
 The operator $d_2$ maps the space of objects like 
 $\cE$  into the space of objects like $\W$, and  $d_2^{\dag}$ maps them the other way. This operator takes the form, in a notation to be introduced below:
\be
d_2 = 
 g_{abc} A^b A^c \oC_{\dot \a}\fr{\pa}{\pa \oy_{a \dot \a}}
+
 \og^{abc} \A_b \A_c C_{ \a} \fr{\pa}{\pa \y^a_{\a}}
 \ee
\item Each of the three pieces of the tripet $(\cC, \cE, \W)$  has the same dimension and quantum numbers, except that $\cC$ has ghost charge minus one, $\cE$ has ghost charge zero and $\W$ has ghost charge one.  
These are derived in section \ref{nightmare} below.
The operators $d_2$  map the space of objects like 
 $\cE$  into the space of objects like $\W$, and  $d_2^{\dag}$ maps them the other way.  The operators $d_3$  map the space of objects like 
 $\cC$  into the space of objects like $\cE$, and  $d_3^{\dag}$ maps them the other way. 
\een

\subsection{\Large Plan of this paper}

\refstepcounter{orange}
{ \theorange}.\;In section \ref{onesec}, we start with a general description of the mathematics of spectral sequences.  These are used in the rest of the paper. Then we introduce the Wess Zumino action, in the BRS form as suggested by Zinn-Justin, then the master equation,  and  nilpotent BRS operator $\d_{\rm BRS}$ that follow from that action. Next the Grading is chosen.  

\refstepcounter{orange}
{ \theorange}.\;
This is a very important choice, and it completely determines and generates the particular spectral sequence that we use in this paper. The spectral sequence  consists of a series of subspaces $E_{\infty}\subset\cdots E_{r+1}\subset E_r \cdots \subset E_0$ and nilpotent differentials $d_r$ defined on each of those subspaces.  The space $E_{\infty}$ is computed by successive approximations  and it is isomophic to the cohomology space of interest, which is $\cH=\ker\d_{\rm BRS}\cap \ker\d_{\rm BRS}^{\dag}$.  A positive definite metric is introduced to define the adjoints $\d_{\rm BRS}^{\dag}$ and $d_{r}^{\dag}$.  The space $E_{\infty}$ is very simple compared to 
$\cH$ and we can find the form of $\cH$ by computing $E_{\infty}$. The BRS cohomology space $\cH$ tells us about the invariants and the anomalies of the theory. In particular we can find the exotic sets $ (\cC,\cE, \W) \subset E_{\infty}$, and from that we can deduce the form of their  counterparts $(\cC,\cE, \W)\subset \cH$.

\refstepcounter{orange}
{ \theorange}.\;In section \ref{Spssection} we review the details of the spectral sequence and define the Elizabethan drama.  In section \ref{simpsec}  the Elizabethan drama is illustrated for some low dimensional spin zero pieces of the cohomology.  In section  \ref{operatorchapter}  an operator method that gives general results for the C sector and $\oC$ sector, separately,  is derived in detail.  Section  \ref{phiintro}  looks at the fairly simple low dimensional spin zero pieces of the cohomology when it includes a constant spinor  $\f_{\dot\a}$.  Section \ref{dimzerosec} contains the analysis of the dimension zero terms with a $\f_{\dot a}$. This is where  the new exotic pair $ (\cE, \W)$  results are derived. 
Section \ref{nightmare} contains the analysis of the dimension one terms with a $\f_{\dot \a}$. This is where  the new exotic triplet $ (\cC,\cE, \W)$ results are derived. 
  Section \ref{sumsec} is a summary of this spectral sequence method for the SUSY WZ model.  Section \ref{conclussec} is the conclusion.  Then in section  \ref{glossary} there is a glossary for the benefit of readers.

\subsection{\Large  Some General Remarks about spectral sequences}
\la{quotesaboutspecseq}
 This paper uses  spectral sequences \ci{dixspecseq}  to deal with the BRS cohomology, and it uses the results in  \ci{dixmin} as an essential part of that analysis.  In addition, it uses the technique known as the \ED.  These are methods well known in mathematical topology circles, but they are unfamiliar to most physicists.  So here are some useful  quotes that talk about \Sps s from the point of view of mathematicians:

``... the behaviour of this spectral sequence...is a bit like an Elizabethan drama, full of action, in which the business of each character is to kill at least one other character, so that at the end of the play one has the stage strewn with corpses and only one actor alive (namely the one who has to speak the last few lines).''  J. F. Adams, as quoted in  \ci{mccleary}, chapter 6.

``A spectral sequence is an algebraic object, like an exact sequence, but more complicated''  J. F. Adams, as quoted in  \ci{mccleary}, chapter 2.

``The machinery of spectral sequences, stemming from the algebraic work of Lyndon and Koszul, seemed complicated and obscure to many topologists. Nevertheless, it was successful...'' G .W Whitehead, quoted in  \ci{mccleary}, chapter 3.

``Topologists commonly refer to this apparatus as `machinery'. ''  J. F. Adams, as quoted in  \ci{mccleary}, chapter 11.

{

{

\subsection{\Large    From Topology to BRS Cohomology in Quantum Field Theory:} 
\refstepcounter{orange}
{ \theorange}.\; This paper will use the `machinery' of spectral sequences for the BRS cohomology of SUSY.  The above quotes are certainly applicable to the work we need to do here.  However the machinery used here is not very similar to that discussed in \ci{mccleary}.  Our machinery works in an inner product space with Fock creation and annihilation operators made from the fields and their derivatives.  Here is what the textbook  \ci{mccleary} says about the techniques that will be used in this paper:

``The BRS operator determines a differential on the Fock space of integrated local polynomial functions of a Yang-Mills field and a Faddeev-Popov ghost field. The resulting cohomology determines invariants of a gauge system, such as the ghost numbers, the Lorentz character and discrete symmetries.   \ci{dixspecseq} filtered the space on which the BRS operator acts and deduced the associated spectral sequence.  The induced grading from the $E_{\infty}-$ term of the spectral sequence decomposes the resulting cohomology in simpler pieces that are computable.'' \ci{mccleary} p. 520.

{
\subsection{\Large    BRS cohomology of SUSY using spectral sequences:} 
\refstepcounter{orange}
{ \theorange}.\; The techniques in \ci{dixspecseq} were applied  for supersymmetric theories in  \ci{dixmin,holes, holescommun,dixminram}.    The papers \ci{dixmin,holes, holescommun,dixminram} left some serious questions unanswered. In fact it was not easy to proceed in a general way, because a good choice of Grading was not obvious, and it had not been found. 

\refstepcounter{orange}
{ \theorange}.\;However, recently, a new and better Grading and a concentrated effort at the \ED\ have been succesful.
The results were summarized in the short paper \ci{dixexotic}. 
This paper will expand upon, and review, and demonstrate, those results with more detail. In addition, the more careful, and more detailed, derivation here, results in some corrections to that paper, notably to the discovery of the exotic triplet.

\refstepcounter{orange}
{ \theorange}.\;{   The above quotes}  about the behaviour of the spectral sequence, are quite applicable to our situation.  The power of the spectral sequence to do the BRS cohomology here necessarily brings in complications, obscurity, and Elizabethan drama-like features\footnote{See the remarks about the \ED\  in the Glossary for further information.}. The length and complications in this paper are proof of those features.

 \subsection{\Large The Wess Zumino SUSY Action and its BRS operator}

\refstepcounter{orange}
{ \theorange}.\;  Here is the action for the Wess Zumino (WZ)   model,  with masses and interactions, including the pseudofields\footnote{See the Glossary for terms like pseudofields.} that are needed to formulate the BRS identity and the master equation of the theory.
\be
\cA= \cA_{\rm WZ}+ \cA_{\rm Structure}
\ee
where
\be
\cA_{\rm Structure}=  X_{\a \dot \b} C^{\a} \overline{C}^{\dot{\b}}
\la{structureaction}\ee
and
\begin{equation} 
\la{actionWZ}
\cA_{\rm WZ} = -
\int d^{4}x \lt \{
\partial_{\mu} A^a \partial^{\mu} {\ov A}_a
+ \y^{a \a}\pa_{\a \dot \b}  \ov{\y}_a^{\dot \b} - F^a \ov{ F}_a 
\ebp
+
m^2 g_{a}
 F^a
+
m g_{(ab)}
\lt ( 
A^a F^b
-
\fr{1}{2}\y^{a \b}  
\y^{b}_{ \b}  
\rt )
+ g_{(abc)} \lt (
F^a A^b A^c -  \y^{a\a} \y^b_{\a} A^c \rt )\ebp
+
\om^2 \og^{a}
 \oF_{a} 
+
\om \og^{(ab)}
\lt ( \A_a \oF_{b} 
-
\fr{1}{2}
\oy_a^{\dot \b}  
\oy_{b,\dot \b}  
\rt )
+ \overline{g}_{abc} \overline{F}_a \overline{ A}_b \overline{ A}_c - \overline{g}^{abc} \overline{\y}_a^{\dot \a} \overline {\y}_{b \dot \a} 
\overline{ A}_c
\ebp
+ \Lam_a  \lt ( \oC^{\dot \b} \pa_{\a \dot \b} \y^{a \a}+
\x^{\mu} \partial_{\mu} F^a 
\rt ) +  \G_a  \lt ( C^{\a} \y^{a}_{\a}+
\x^{\mu} \partial_{\mu} A^a 
\rt )
\ebp
+ Y_a^{\a}
\lt (  \pa_{\a \dot \b}   A^a
\overline{C}^{\dot{\b}}
+ F^a C_{\a}
+ \x^{\mu} \partial_{\mu} \y^a_{\a} \rt ) 
+ \ov\Lam^a  \lt ( C^{\a} \pa_{\a \dot \a} \oy_{a}^{ \dot \a}+
\x^{\mu} \partial_{\mu} \oF_a 
\rt )
\ebp+ \ov\G^a  \lt ( \oC^{\dot \a} \oy_{a \dot \a}+
\x^{\mu} \partial_{\mu} \A_a 
\rt ) 
+ \oY^{a \dot \a}
\lt (  \pa_{\a \dot \a}   \A_a 
 C^{\a}
+ \oF_a \oC_{\dot a}
+ \x^{\mu} \partial_{\mu} \oy_a^{ \dot\a} \rt ) 
\rt \}
\ee
\refstepcounter{orange}
{ \theorange}.\;In the above, the scalar field $A^a$, the Weyl spinor field $\y^a_{\a},( \a = 1, 2)$
and the auxiliary field $F^a$ were all contained in the original paper of Wess and Zumino \ci{WZ}.  The complex notation is more useful for our purposes, because the Wess Zumino model is naturally complex. Their complex conjugates are  $\A_a; \oy_{a\dot\a}, (\dot \a = \dot 1, \dot 2); \ov F_a$.  The index $a= 1\cdots n$ labels the set of fields, and it might include isospin or some sort of representation of a gauge group. 
The pseudofields $\G_a, Y_{a}^{ \a}, \Lam_a$ are sources for the BRS variations of the three fields $A^a,\y^a_{\a},F^a $, respectively.  Their complex conjugates are  $\ov \G^a, \oY^a_{\dot \a}, \ov \Lam^a$.  The BRS transformations are characterized by the spacetime constant Grassmann even spinor ghost field $C_{\a}$ and its complex conjugate $\oC_{\dot\a}$.  We need to close the algebra with a Grassmann odd exterior derivative $\x^{\a \dot \a}\pa_{\a \dot \a}$.

\refstepcounter{orange}
{ \theorange}.\;{ The master equation (Zinn-Justin identity)} summarizes the invariance of the action and the nilpotence of the variations of the fields:
\be
\int d^4 x \lt \{
\fr{\d \cA}{\d A^a} \fr{\d \cA}{\d \G_a}
+
\fr{\d \cA}{\d \y^{a\a}} \fr{\d \cA}{\d Y_{a\a}}
+
\fr{\d \cA}{\d F^a} \fr{\d \cA}{\d \Lam_a}
+
\fr{\d \cA}{\d \A_a} \fr{\d \cA}{\d \ov\G^a}
\ebp+
\fr{\d \cA}{\d \oy_a^{\dot \a}} \fr{\d \cA}{\d \oY^a_{\dot \a}}
+
\fr{\d \cA}{\d \oF_a} \fr{\d \cA}{\d \ov\Lam^a}
\rt \} + 
\fr{\pa \cA}{\pa X_{\a \dot \b}} 
\fr{\pa \cA}{\pa \x^{\a \dot \b}}   =0
\la{mastercss}
\ee
This equation gives rise to a nilpotent `square root' operator $\d_{\rm BRS}\equiv \d$. These are the BRS transformations of the fields and the pseudofields, as follows:

\be
\la{brstransWZ} 
\vspace{.1in}
\framebox{{$\begin{array}{lll}  
\\ & &{\rm Nilpotent  \;Transformations \;\d \equiv \d_{\rm BRS}\;including\;pseudofields}\\
\d A^i&= & 
\fr{\d {\cal A}}{\d \G_i} 
=  \y^{i}_{  \b} {C}^{  \b} 
+ \x^{\g \dot \d} \partial_{\g \dot \d} A^i
\\
\d {\ov A}_i&= & 
\fr{\d {\cal A}}{\d {\ov \G}^i} 
=  {\ov \y}_{i  \dot \b} {\ov C}^{ \dot  \b} 
+ \x^{\g \dot \d} \partial_{\g \dot \d} {\ov A}_i
\\

\d \y_{\a}^i &  =& \fr{\d {\cal A}}{\d {  Y}_i^{   \a} } = 
\pa_{ \a \dot \b }  A^{i} {\ov C}^{\dot \b}  
+ 
C_{\a}   
F^i
+ \x^{\g \dot \d} \partial_{\g \dot \d}  \y^{i}_{\a  }
\\

\d
 {\ov \y}_{i \dot \a} &  =& 
\fr{\d {\cal A}}{\d { {\ov Y}}^{i \dot   \a} } = 
\pa_{ \a \dot \a }  {\ov A}_{i} {C}^{\a}  
+ 
{\ov C}_{\dot \a}   
{\ov F}_{i}
+ \x^{\g \dot \d} \partial_{\g \dot \d} 
 {\ov \y}_{i \dot \a} 
\\
 \d F^i 
&=&\fr{\d {\cal A}}{\d \Lam_i} = 
  \pa_{\a \dot \b}   \y^{i \a} {\ov C}^{\dot \b} 
+ \x^{\g \dot \d} \partial_{\g \dot \d}  F^i 
\\
 \d \oF_i 
&=&\fr{\d {\cal A}}{\d \ov\Lam^i} = 
  \pa_{\a \dot \b}   \oy_{i}^{\dot \a} { C}^{\b} 
+ \x^{\g \dot \d} \partial_{\g \dot \d}  \oF_i 
\\
\d \G_i 
&= &
 \fr{\d {\cal A}}{\d A^i} 
=
  \Box    {\ov  A}_{i} 
  +   m {g}_{iq} F^q  +2 g_{ijk} ({ A}^{j}{ F}^{k}- {\y}^{j\a}{\y}_{\a}^{k})
-\pa_{ \a \dot \b } Y_{i}^{ \a}    {\ov C}^{\dot \b}   
+ \x^{\g \dot \d} \partial_{\g \dot \d} \G_i
\\
\d {\ov \G}^i 
&= & \fr{\d {\cal A}}{\d {\ov A}_i} 
=
 \Box          { A}^{i} 
+  m {\ov g}^{ij} {\ov  F}_{ j}  
+ 2 {\ov g}^{ijk}      ({ \A}_{j}{ \oF}_{k}- {\oy}_{j}^{\dot \a}{\oy}_{k \dot \a})
- \pa_{ \a \dot \b } {\ov Y}^{ i \dot \b}    {C}^{\a}   
+ \x^{\g \dot \d} \partial_{\g \dot \d} 
 {\ov \G}^i
\\
\d Y_{i}^{ \a} 
&=&\fr{\d {\cal A}}{\d {  \y}^i_{   \a}} 
= 
-
  \pa^{\a \dot \b  }   
{\ov \y}_{i   \dot \b}
+  m {g}_{iq}   
\y^{q \a} 
 +
2 g_{ijk}  \y^{j \a} A^k    
-
\G_i  
 {C}^{  \a}
+ 
  \pa_{\a \dot \b}  \Lam_i   {\ov C}^{\dot \b} 
  + \x^{\g \dot \d} \partial_{\g \dot \d}  Y_{i}^{ \a}
\\
\d 
{\ov Y}^{i \dot \a} 
&=&\fr{\d {\cal A}}{\d {\ov \y}_i^{ \dot \a} 
} 
= 
-
  \pa^{\b \dot \a  }   
{ \y}^i_{ \b}
+  m {\ov g}^{ik}   
{\ov \y}_{k}^{\dot  \a} 
+
2 {\ov g}^{ijk} {\ov \y}_{j}^{\dot  \a} 
{\ov A}_k  
-
{\ov \G}^i  
 {\ov C}^{\dot  \a}
+ 
  \pa^{\b \dot \a}  \ov\Lam^i   { C}_{\b} 
+ \x^{\g \dot \d} \partial_{\g \dot \d}  
{\ov Y}^{i \dot \a} 
\\
 \d \Lam_i  
&=&  \fr{\d {\cal A}}{\d {F}^i } =
\oF_i + m g_{ib}  A^b+  g_{ibc}  A^b A^c
+ Y_i^{\a}
  C_{\a}+ 
\x^{\g \dot \d} \partial_{\g \dot \d}   \Lam_i  
\\
 \d {\ov \Lam}^i  
&=&  \fr{\d {\cal A}}{\d {\ov F}_i } =
F^i +m \og^{ij}   \A_j + \og^{ijk}  \A_j \A_k
+ \oY^{i \dot \a}
  \oC_{\dot\a}+ 
\x^{\g \dot \d} \partial_{\g \dot \d}   {\ov \Lam}^i   
\\

\d \x_{\a \dot \b} 
&=& \fr{\pa {\cal A}}{\pa { X}^{\a \dot    \b}} 
=
 -   C_{\a} {\ov C}_{\dot \b}
\\
\d X_{\a \dot \b} 
&=& \fr{\pa {\cal A}}{\pa { \x}^{\a \dot    \b}} 
= 
\int d^4 x \; \X_{\a \dot \b} 
\\
\d C_{\a}
&=&
0
\\
\d  {\ov C}_{\dot \b}
&=&
0
\\
\end{array}$}} 
\ee
\Large
 \section{\LARGE   The  spectral sequence: Grading, Counting Operators and the Operators $\d_r, r=0,1,2$  the space $E_1$ and the Elizabethan drama}
\la{Spssection}

The spectral sequence for an operator is entirely generated by the  grading we choose. Here we choose:
\la{countingops}
\be
N_{\rm Grading}
=
N_{\rm C} 
+
N_{\rm \oC} 
+ 
2 N_{\x}+
N_{\rm m} 
\eb
+ N_A  
+N_{\y} +N_{F} 
+    N_{\G}  +N_{Y}
   +N_{\Lam}  
\eb
+ N_{\A}  
+N_{\oy} +N_{\oF} 
     +
    N_{\ov\G}  +N_{\ov Y}
   +N_{\ov\Lam}  
   \la{Ngrading}
   \ee

The dimension is
\be
{\rm   Dim }  =N_{\rm Derivatives}   +{\rm    Dim}_{\rm fields \; etc.}
\ee
where
\be
N_{\rm Derivatives}    ={\rm Count\; the\;  derivatives\;\pa}
\la{dimofderivatives}
\ee
and
\be
{\rm    Dim}_{\rm fields \; etc.}=+
 N_{\rm m} 
 \eb
 +  N_A  
+\fr{3}{2}  N_{\y} + 2 N_{F} 
 +
 N_{\A}  
+\fr{3}{2}  N_{\oy} + 2 N_{\oF} 
\eb + 3 N_{\G}   + \fr{5}{2} N_Y + 2 N_{\Lam}    
+ 3 N_{\ov\G}   + \fr{5}{2} N_{\ov Y} + 2 N_{\ov \Lam}    
\eb  - \fr{1}{2} N_C- \fr{1}{2} N_{\oC}  
-  N_{\x} +   N_{X}
\ee
The ghost charge is:
\be
N_{\rm ghost}= -4 + N_C+  N_{\oC}  
+  N_{\x} - 2    N_{X}
- N_{\G}   - N_Y -  N_{\Lam}    
-N_{\ov\G}  - N_{\ov Y} - N_{\ov \Lam}    
\ee
The form charge is:
\be
N_{\rm form}=
N_{\rm ghost}+ 4
\ee

\subsection{\Large  Important Commutators:} We note that 
\be
\d = \d_{\rm BRS} =\d_{\rm WZ} + \d_{\rm Structure}
\ee
\be
\cA = \cA_{\rm BRS} =\cA_{\rm WZ} + \cA_{\rm Structure}
\ee
and
\be
\lt [ {\rm Dim }, \d_{\rm BRS} 
\rt ] =0
\ee
\be
\lt [  {\rm Dim }, \cA_{\rm BRS} 
\rt ] =0
\ee
\be
\lt [ N_{\rm ghost}, \d_{\rm BRS} 
\rt ] =\d_{\rm BRS} 
\ee
\be
\lt [ N_{\rm ghost}, \d_{\rm WZ}
\rt ] =\d_{\rm WZ}
\ee
 
\be
\lt [ N_{\rm ghost}, \d_{\rm Structure}
\rt ] =\d_{\rm Structure}
\ee
\be
\lt [ N_{\rm ghost}, \cA_{\rm BRS}
\rt ] =0
\ee\be
\lt [ N_{\rm ghost}, \cA_{\rm WZ}
\rt ] =0
\ee
 \be
\lt [ N_{\rm ghost}, \cA_{\rm Structure}
\rt ] =0
\ee

\la{summaryofdeltar}
 
\refstepcounter{orange}
{ \theorange}.\;{ The reason for this choice of grading} is that:
\ben
\item
 The form of $E_1$ is easy, and we eliminate all the pseudofields from the spaces 
$E_r, r\geq1$ of the spectral sequence this way.
\item
At the same time, by including the terms $N_{\rm C} 
+
N_{\rm \oC} 
+ 
2 N_{\x}$, we get to include the very important results for the operator $\d_{\rm Structure}= C   \s^{\m}  \overline{C}   \x^{\mu \dag} \equiv C_{\a} \oC_{\dot \b}  \x_{\a \dot \b}^{ \dag}    $.\een
  This means that these pseudofields do not appear in any of the spaces $E_r, r\geq 1$, since they are all subspaces
\be
 E_{\infty} \subset \cdots E_{r+1}\subset E_{r } \subset \cdots E_{0}, r = 1,2,\cdots 
\ee
where our initial space is  $E_{0}$, and where the BRS operator $\d$ is given by the table (\ref{brstransWZ}). Also the structure of the ghosts in $E_r, r \geq 1$ is fixed from the beginning, which turns out to be very useful. It is a bit complicated, but so is the result for $E_{\infty} \approx \cH$.

\refstepcounter{orange}
{ \theorange}.\;
This does not necessarily mean that the pseudofields do not appear in the final cohomology space $\cH \approx E_{\infty}$, however, as we shall see. Also the ghost structure in the space $\cH \approx E_{\infty}$ changes from $E_{\infty}$ too.  The pseudofields and the ghosts  are linked, of course.

\refstepcounter{orange}
{ \theorange}.\;This grading determines that the graded operators are as follows:
\be  \d= \sum_{r=0}^2 \d_r \; {\rm where}
;\;  [ N , \d_r]=r\d_r
\ee and
\be \d_0 = \d_{0,\rm WZ} + \d_{\rm Structure} 
\la{deltazero}
\ee
where
\be
\d_{\rm Structure} = C_{\a} \oC_{\dot \b}  \x_{\a \dot \b}^{ \dag}  
\la{deltastructure}
\ee
and
\be \d_{0,\rm WZ} = \int d^4 x \lt \{ \Box \A_a \fr{\d}{\d \G_a}
+\Box A^a \fr{\d}{\d \oG^a} 
+
\pa_{\b \dot \b} \oy_a^{\dot\b} \fr{\d}{\d Y_{a\b}}
\ebp+  
\pa_{\a \dot \b} \y^{a\a} \fr{\d}{\d \oY^a_{\dot \b}}
+  \oF_a \fr{\d}{\d \Lam_a}
+F^a \fr{\d}{\d \ov \Lam^a}\rt \}
\la{deltazeroWZ}\
\ee
and
\be
\la{deltaoneWZ}
 \d_{1,\rm WZ} = 
  C^ {\a} \na^{\rm Big}_{\a} + \oC^{\dot \a} \ov\na^{\rm Big}_{\dot\a} + \d_{\rm  EM}
\ee
\refstepcounter{orange}
{ \theorange}.\;We will see that only a small part of the first two operators in (\ref{deltaoneWZ}) is actually needed for most of the spectral sequence. We denote the big operators with the word `Big' as in  $\na^{\rm Big}_{\a}$, and we will use much smaller versions   $\na_{\a}$, without the word Big,    in  section \ref{nanoprime}.  But the full operators $\na^{\rm Big}_{\a}$ will also be needed to see that the full operator is actually nilpotent, and some of its  terms will also be needed to construct the higher terms in the spectral sequence. The full operators are as follows:
\be
 C^ {\a} \na^{\rm Big}_{\a} = \int d^4 x \lt \{ C^{\a} \y^i_{\a}  \fr{\d}{\d A^i}
 +
  \pa_{\a \dot \b}
   \A_a C^{\a}  
\fr{\d}{\d \oy_{a\dot \b}}
+
 F^a C_{\a}  \fr{\d}{\d \y^a_{\a}}
+  C^{\a} \pa_{\a \dot \a} \oy_{a}^{ \dot \a} 
 \fr{\d}{\d \oF_a}
\ebp
+
 \lt (Y_a^{\a} C_{\a}   \rt ) \fr{\d}{\d \Lam_a}
+\lt ( 
\G_a C_{\a} \rt ) \fr{\d}{\d Y_{a \a}}
+ \lt ( \pa_{\a \dot \a} \ov\Lam^a C^{\a} 
\rt ) \fr{\d}{\d \oY^a_{\dot\a}} 
+ \lt (   \pa_{\a \dot \b} \oY^{a \dot \b} C^{\a}   \rt ) \fr{\d}{\d \oG^a}
\rt \}
\la{naprime}\ee
and
\be
\oC^{\dot \a} \ov\na^{\rm Big}_{\dot\a} 
= \int d^4 x \lt \{\oC \cdot \oy_{i} \fr{\d}{\d \A_i}
+ \pa_{\a \dot \b}
   A^a \oC^{\dot \b} \fr{\d}{\d \y^a_{\a}} 
   \pa_{\a \dot \b}
   + \ov F_a \oC_{\dot \b}  
\fr{\d}{\d \oy_{a\dot \b}}+
  \oC^{\dot \b} \pa_{\a \dot \b} \y^{a \a} 
   \fr{\d}{\d F^a}
\ebp
   +\lt ( \oY^{a \dot\a} \oC_{\dot \a}    \rt )  \fr{\d}{\d \ov \Lam^a}
+\lt ( \pa_{\a \dot \b}\Lam_a \oC^{\dot \b} \rt ) \fr{\d}{\d Y_{a \a}}
+ \lt (   \ov \G^a \oC_{\dot \a}  
\rt ) \fr{\d}{\d \oY^a_{\dot\a}}
+ \lt (   \pa_{\a \dot \b} Y_a^{\a} \oC^{\dot \b}   \rt ) \fr{\d}{\d \G_a}
\rt \}
\la{naprimebar}\ee 
\refstepcounter{orange}
{ \theorange}.\;The equation of Motion terms are defined by:
\be
\d_{\rm  EM}=\int d^4 x \lt \{
 \lt (   g_{a}  m^2 +  g_{ab}  m  A^b+  g_{abc}  A^b A^c   \rt ) \fr{\d}{\d \Lam_a}
\ebp+\lt (  \og^{a} m^2  +\og^{ab} m  \A_b + \og^{abc}  \A_b \A_c   \rt )  \fr{\d}{\d \ov \Lam^a}
\la{thisone1}
\ebp 
+\lt ( -  g_{ab} m \y^b_{\a}  -2   g_{abc} A^b \y^c_{\a} \rt ) \fr{\d}{\d Y_{a \a}}
\ebp
+ \lt ( -  \og^{ab} m \oy_{b\dot \a}- 2\og^{abc} \A_b \oy_{c\dot \a}
\rt ) \fr{\d}{\d \oY^a_{\dot\a}}
+\ebp+  \lt (  g_{abc } \y^{b\a} \y^c_{\a} +  m  g_{ab } F^b+ 2 g_{abc } A^b F^c\rt ) \fr{\d}{\d \G_a}
\ebp
+ \lt (\og^{abc} \oy_b \cdot\oy_c + m \og^{ab} \oF_b+2 \og^{abc} \A_b   \oF_c \rt ) \fr{\d}{\d \oG^a}
\rt \}
\la{eqmoterms}
\ee
and the Exterior Derivative term is: 
\be
\d_2 =  \x^{\mu} \partial_{\mu}   
\la{extderd2}
\ee

\subsection{\Large The Space $E_1$ and the operator $d_0=\d_0$}
\refstepcounter{orange}
{ \theorange}.\;We need to assume some of the results in \ci{dixspecseq} and \ci{dixmin} for some of the following discussion.    But we will try to recall the major results as we go, and refer the reader to the glossary for further help.
First we note from \ci{dixspecseq} that the space $E_1$ is the kernel of the following (in the notation of \ci{dixspecseq}):
\be
E_1 = \ker (\d_0+\d_0^{\dag})^2\equiv \ker \D_0 = \ker \d_0 \cap \ker \d_0^{\dag} 
\ee
From line (\ref{deltazero}) we see that
\be
\d_0 = \d_{\rm Structure}   +  \d_{0,\rm WZ}
\ee 
where 
 $\d_{\rm Structure}$ is in (\ref{deltastructure}), and
$\d_{0,\rm WZ}$ is in (\ref{deltazeroWZ}).

\refstepcounter{orange}
{ \theorange}.\;  Note that the operators   $\d_{\rm Structure}$  in 
(\ref{deltastructure}) and  
$\d_{0,\rm WZ}$  in (\ref{deltazeroWZ})  are  completely independent of each other, so they, and their adjoints, clearly anticommute, with each other.  So we can write 
 \be
E_1 = \ker \D_{\rm Structure} 
\cap \ker \D_{0,\rm WZ}
\ee

\subsection{\Large  \;Form of $E_1$:}
\la{formofE1para}
\refstepcounter{orange}
{ \theorange}.\;Using the results of \ci{dixmin} for the operator $\d_{\rm Structure}$ and the results of \ci{dixspecseq} for   operators of the form $\d_{0,\rm WZ}$, we easily get the following form, where we show the explicit form in terms of $\x$ and $C$ and $\oC$:  
\be
E_1=M_0 + (C \x \oC) R_0 + {\rm C\;terms}+ {\rm \oC\; terms}
\la{structureofE1line1}
\ee
The  ${\rm C\;terms}$ contain only the ghost C and never the ghost $\oC$:
\[
 {\rm C\;terms}=
 \S_{n=1}^{\infty} P_n  C_{\a_1} \cdots C_{\a_n}
 + \S_{n=0}^{\infty} Q_n  (\x C)_{\dot \a}   C_{\a_1} \cdots C_{\a_n}
\]\be
 + \S_{n=0}^{\infty}(C \x^2 C) T_n C_{\a_1} \cdots C_{\a_n}
\la{structureofE1line2}
\ee
The  ${\rm \oC\;terms}$ contain only the ghost $\oC$ and never the ghost $C$:
 \[
{\rm \oC\; terms}=\S_{n=1}^{\infty} \oP_n   \oC_{\dot \a_1} \cdots \oC_{\dot \a_n}
 +\S_{n=0}^{\infty} \oQ_n (\x \oC)_{ \a}   \oC_{\dot \a_1} \cdots \oC_{\dot \a_n}
\]\be +\S_{n=0}^{\infty}(\oC \x^2 \oC) \oT_n \oC_{\dot \a_1} \cdots \oC_{\dot \a_n}
\la{structureofE1line3}
\ee

\refstepcounter{orange}
{ \theorange}.\;This division into ${\rm C\;terms}$ and  ${\rm \oC\;terms}$ is a very useful  simplification for the spectral sequence, because it allows us, sometimes,  to ignore half of the operator $d_1$ for each part, and that enables us to solve the problem easily.  The first terms in line (\ref{structureofE1line1}) require some extra attention however, as we will see.

\refstepcounter{orange}
{ \theorange}.\;  This expression needs a considerable amount of explanation. 
As a start at that explanation, consider the following term which is at the beginning of equation (\ref{structureofE1line2}) in the above:
\be
 \S_{n=1}^{\infty} P_n  C_{\a_1} \cdots C_{\a_n}
\ee 
 As far as the cohomology goes, at this stage, there is no requirement that the indices $  {\a_1} \cdots  {\a_n}$ on the ghosts $ C_{\a_1} \cdots C_{\a_n}$ be contracted with anything.  So we show them to be uncontracted.  However in the term 
 \be
  \S_{n=0}^{\infty}(C \x^2 C) T_n C_{\a_1} \cdots C_{\a_n}
 \ee
 the expression
 \be
 (C \x^2 C)\equiv C^{\a}\x_{\a\dot \b} \x^{\g\dot \b} C_{\g}
 \ee
 must have the contractions shown--there are no uncontracted indices in it. 
 
 \refstepcounter{orange}
{ \theorange}.\; The explanation of the rather bizarre and complicated form of the expression in terms of the variables $\x,C,\oC$ and their indices is in \ci{dixmin}.  This is much more useful than it might appear to be, since we will use it as the basis of all our further discussions about the BRS cohomology of SUSY.  For Lie groups the analogous situation is much more simple\footnote{The analogy for Lie groups is discussed briefly in \ci{dixmin}.}, but equally important.

\subsection{\Large  More about the form of $E_1$}

\refstepcounter{orange}
{ \theorange}.\; In the above,  all dependence on the ghosts $C,\oC,\x$ is shown explicitly.  The functions $M_0,R_0,P_n,Q_n,T_n$  do not contain any implicit dependence on those ghosts $C,\oC,\x$ or on the pseudofield\footnote{This source plays very little role in the cohomology--it is there just to make the master equation symmetrical. For supergravity it is more important.} $X_{\a \dot \b}$ in line (\ref{structureaction}).   $M_0$ and $R_0$ are real,  and $P_n,Q_n$ and $T_n$ are complex. These coefficients are all polynomial functions of the mass m and the field variables $A^a,\y^a_{\a} $, their complex conjugates $\A_a,\oy_{a\dot\a} $ and the derivative operator $\pa_{\a \dot \b}$, but they do not contain any dependence on  the ghosts $C,\oC,\x,X$ or the pseudofields or the auxiliary F.

\refstepcounter{orange}
{ \theorange}.\;
{ Because of the simple form of the Laplacian operator }$\D_{0,\rm WZ}$ as discussed in \ci{dixspecseq}, it follows that
\bitem
\item
 {{the auxiliary variables}} $F^a,\oF^a $, and 
 \item
 {{all of the pseudofield variables}}
 $\Lam_a, Y_{a\a}, \G_a,  $, and
 \item
 {{all of the complex conjugate pseudofield variables}}  \\
$\ov \Lam^a, \oY^a_{\dot\a}, \ov \G^a, $, and
\item
{{  all of the derivatives of these variables}}, 
\eitem
{{ have all been eliminated from the converging set of spaces $E_r, r\geq 1$. }} The way this works is very simple. 
Much more of this kind of analysis can be found in the references 
\ci{ dixspecseq,dixmin} 
and also \ci{holes, holescommun,dixminram}.

\refstepcounter{orange}
{ \theorange}.\;{  For example} let us consider the piece
\be
\d_{0,{\rm piece}}=  \int d^4 x \lt \{ \Box \A \fr{\d}{\d \G}
 \rt \}
\ee
In the notation of \ci{dixspecseq}, we can write this as
\be
 \int d^4 x   \Box \A \fr{\d}{\d \G}\equiv
\S_{n=0}^{\infty} \h^{\m\n}
\A_{\m \n \n_1 \cdots \n_n }
\G^{\dag}_{ \n_1 \cdots \n_n }
=
\A_{\m}^{\;\; \m}
\G^{\dag} 
+
\h^{\m\n} \A_{\m \n \n_1}
\G^{\dag}_{ \n_1 }
+ \cdots
\ee
Note that the terms $\A, \A_{\m}$ with no derivatives and one derivative do not appear in this sum. When we apply the reasoning in \ci{dixspecseq} here we see that the space $E_1$ can have an arbitrary dependence on these variables $\A, \A_{\m}$. The dependence on $\A$ with two or more derivatives is more complicated, and is treated below. We also see   that all dependence on $\G$ is eliminated from $E_1$ by noting that $\D_0 E_1=0$, since  $\D_0$ is a positive semidefinite sum of terms including terms like $\G_a \G_{a}^{ \dag}+\G_{a\a \dot \b} \G_{a \a \dot \b}^{ \dag}+ \cdots$.

\subsection{\Large Symmetrized Field Variables generate the Space $E_1$}

\refstepcounter{orange}
{ \theorange}.\;We need to specify the field variables that can appear in $E_1$.  The equations are
\be 
A _{\a \dot \b, \g \dot \d}^{a \dag} \ve_{\a \g}\ve_{\dot \b \dot \d}E_1=0
\ee
We use the notation
\be
\Box A =  \h^{\m\n} \pa_{\m}\pa_{\n}A^a= A^{a  }_{\m\n} \h^{\m\n} \equiv A _{\a \dot \b, \g \dot \d}^{a } \ve^{\a \g}\ve^{\dot \b \dot \d}
\ee
We can define
\be
 A _{\a \dot \b, \g \dot \d}^{a } 
\equiv A _{(\a \g), ( \dot \b \dot \d)}^{a}
+A _{[\a \g],  [\dot \b \dot \d]}^{a} 
\ee
where
\be
A _{(\a \g), (\dot \b \dot \d)}^{a}
= \fr{1}{2} \lt (
A _{\a \g,  \dot \b \dot \d}^{a} 
+
A _{\g \a,  \dot \b \dot \d}^{a} 
\rt)
\ee
\be
A _{[\a \g],  [\dot \b \dot \d]}^{a}
= \fr{1}{2} \lt (
A _{\a \g,  \dot \b \dot \d}^{a} 
-
A _{\g \a,  \dot \b \dot \d}^{a} 
\rt)
\ee
This uses the fact that \be
A _{\a \dot \b, \g \dot \d}^{a }=
A _{ \g \dot \d,\a \dot \b}^{a }
\ee
which is equivalent to 
\be
A^{a  }_{\m\n} 
=A^{a  }_{\n\m} = \pa_{\m}\pa_{\n}A^a
\ee 
We see that in fact there are really  only two independent variables here, namely the totally symmetrized variable
$A _{(\a \g),  (\dot \b \dot \d)}^{a} $  and the totally antisymmetrized variable $A _{[\a \g],  [\dot \b \dot \d]}^{a}$. This kind of reasoning shows very quickly that the only variables that survive to $E_1$ are the totally symmetrized variables, namely
\be
A ^{a };  A _{\a, \dot \b}^{a }; A _{(\a \g),  (\dot \b \dot \d)}^{a}; A _{(\a \g \ve),  (\dot \b \dot \d\dot \z)}^{a} 
\cdots \ee
\be
\y^{a }_{\a} ;  \y^a_{(\a \b), \dot \b} ; \y^a_{(\a \b \g), (\dot \b \dot \d)} ; \y^a_{(\a \b\g \d ) ,(\dot \b \dot \d\dot \z)} 
\cdots \ee
and their complex conjugates
\be
\A _{a };  \A _{a \dot \a, \b} ; \A _{a (\dot \b \dot \d),(\a \g)} ; \A _{a  (\dot \b \dot \d\dot \z),(\a \g \ve),} 
\cdots \ee
\be
\oy_{a  \dot \a} ;  \oy_{a (\dot\a \dot\b), \b} ; \oy_{a  (\dot \b \dot \g \dot \d),(\a \b ), } \cdots \ee
This valuable information,  together with the structure in \ci{dixmin}, will allow  us to completely solve the cohomology using the spectral sequence. 
In this paper we will drop the dependence on the mass parameter m, since it makes everything much longer.  Replacing it is not difficult but it does increase the complications, a lot.

\subsection{\Large The power of this form of $E_1$}
\refstepcounter{orange}
{ \theorange}.\;  All the ghost charge of the space $E_1$ is detemined by the ghosts $\x,C,\oC$ alone since these are the only objects with nonzero ghost charge left in $E_1$, since all of the antiflelds are excluded from $E_1$, and all the fields have zero ghost charge\footnote{We ignore the constant pseudofield $X_{\a \dot \b}$, because it plays no role unless we consider supergravity. }.  As we know from \ci{dixspecseq}, after we have dealt with the issues of exterior derivatives and integration by parts, including terms like $\x \pa$,  the integrated polynomlals, like the action, are all of the form
\be
\int d^4 x \cP[{\rm fields, pseudofields, C ,\oC}] \equiv (\x^4) \cP[{\rm fields, pseudofields, C , \oC}]  
\ee
and so we are particularly interested in ghost  charges -1,0,1.  Ghost charge -1 corresponds to terms like
\be
\int d^4 x \G_i  
\ee
which are in the cohomology space for the free massless theories, ghost charge 0 corresponds to terms like the action itself, and ghost charge 1 corresponds to anomalies.

\subsection{\Large  The operator  $d_1$ and the space $E_{2}$}

 \refstepcounter{orange}
{ \theorange}.\;
The next approximation to the cohomology space is \ci{dixspecseq}:
\be
E_2 = \ker \D_1 = \ker d_1\cap \ker d_1^{\dag}
\la{eqforE2}
\ee
where
\be
d_1 = \P_1 \d_1 \P_1
\ee
where $\P_1$ is the orthogonal projection operator onto the space $E_1$, and where the operator $\d_1$ is in (\ref{deltaoneWZ}).

Sometimes  the best alternative is to use the form 
\be
E_2 =  \ker d_1\cap \ker d_1^{\dag}
\la{eqforE2easier}
\ee
and evaluate it in small pieces, as we shall now proceed to do. This procedure is the Elizabethan drama. 

\subsection{\Large The \ED\ and the operator $d_1$}
\la{ghostquotes}

\refstepcounter{orange}
{ \theorange}.\;Because the spectral sequence consists of a set of subspaces $E_{\infty} \subset\cdots E_r \cdots$ $\subset E_1 \subset E_0$ that are the kernels $E_{r+1} =  ( \ker d_r\cap \ker d_r^{\dag}) \cap E_r$ of nilpotent differentials $d_r$ operating in the subspace $E_{r} $, it follows that whenever we have a mapping $t_1 \in E_r\stackrel{d_r}{\lra}  t_2\in E_r$, it means that both of the terms $t_1 \in E_r$ and  $ t_2\in E_r$ are removed from the space $E_{r+1}$.  This is what is meant by being `killed', as in an Elizabethan drama.

\refstepcounter{orange}
{ \theorange}.\;The work that comes next, as noted by Adams and Whitehead, is  rather complicated and obscure,
and very much like an Elizabethan drama \ci{shake}.  It is an amusing coincidence that ghosts play an important role in the spectral sequence for BRS cohomology. There were no ghosts in the original mathematical literature.   Yet the word ``spectral'' in English primarily means ``of or like a ghost". Probably, this originates in the translation from the original French where `Sequence Spectrale' was probably meant to refer to a sequence of spectra of differential operators.   The material that arises from the \ED\ is indeed a complicated and obscure Eliabethan drama with many bodies.

\subsection{\Large Summary of the Spaces  $E_r, r\geq 2$ for the BRS Cohomology of Pure WZ theory}

\refstepcounter{orange}
{ \theorange}.\;At this level we have derived what we can from $d_1$ and from 
$\D_0$. Let us summarize what we know at this stage: 
\ben
\item
We know that the space $E_1$ and so the       spaces $E_r\subset E_1, r\geq 2$ are functions of the symmetrized fields, without any pseudofields or auxiliaries.
\item
We know that the   space $E_1$ and so the spaces $E_r\subset E_1, r\geq 2$ the ghosts $C,\oC, \x$ appear only in those combinations which constitute the cohomology of the structure cohomology operator $\d$ in (\ref{brstransWZ}).
\item
We know that $d_1$ commutes with $N_{\x}$ so we can look at the different situations for $N_{\x}=0,1,2$ separately at level $E_2$ and $E_1$.
\item
We know that $d_r$ commutes with dimension so we can look at the different eigenspaces that characterize  dimension  separately, at all levels. These go from a small negative integer like -4 to $+ \infty$. 
\item
We will see that the operator $d_2$ takes different forms for different subspaces.  We will look at a few of those forms in this section.

\een

  Now we have had a first glance at $E_1$ for the pure WZ model.  Clearly there are parts with spin up to infinity.

\section{\LARGE Choosing some simple sectors to introduce the Use of the \ED }
\la{simpsec}

\refstepcounter{orange}
{ \theorange}.\;
Let us recall here that the space $E_1$ divides into sectors with $N_{\x}=0,1,2$ and that the operator $d_1$ commutes with $N _{\x}$ and dimension and $\rm N_{m}$.  As a result we can consider subspaces with eigenvalues of  $N _{\x}$ and dimension and $\rm N_{m}$   separately when considering the subspace $E_1$, the action of the operator  $d_1$   and the subspaces of  $E_2$.

 \refstepcounter{orange}
{ \theorange}.\;
From now on, in this paper, we will set the mass m to zero. This avoids quite a lot of complication and extra terms, and it is not hard to replace it once we see how the various terms work.  

We will also concentrate now on terms with spin zero.  We will return to terms with non-zero spin later, but for now this is another useful simplification.  We have some general results about the form of $E_1$ and we will assume that we start in the space $E_1$ from here on.

We will work our way up in dimension for these types of terms. This is a useful exercise in understanding the spectral sequence as it develops here. 
 Remember that the operators $\d, \d_r, d_r, r= 0,1,2\cdots$ all commute with dimension.  

\subsection{\Large Low dimension examples of terms in $E_1$ and $E_2$  and some mappings}

\refstepcounter{orange}
{ \theorange}.\;The terms with zero spin in the sectors $(1,C,C C) $ start with dimension one:
\be
 A^a, \A_a,\y^a_{\a}  C^{\a} ,\oy_{a\dot \a} \oC^{\dot \a} 
\ee
 So we can ignore them to start with.

\refstepcounter{orange}
{ \theorange}.\;The terms with zero spin in the sectors $(C\x) $ start with dimension zero: 
\be
(C\x\oy),
(\oC\x\y)
\ee
So we can ignore them too  to start with.

\refstepcounter{orange}
{ \theorange}.\;The terms with zero spin in the sectors $(C\x\oC) $ start with dimension minus one.  
\be
(C\x\oC) A,
(C\x\oC) \A,
\ee
So we can ignore them too  to start with.

\subsection{\Large The  Sector for dimension =-2 and spin =0}
\refstepcounter{orange}
{ \theorange}.\;
The lowest possible dimension for spin zero is -2, so we will look at that.
Here are all the possible terms with dimension =-2 and spin =0 in the Space $E_1$:
\be
    e^{a} \Ct   \A_a , 
    e_{a} \Ct   A^a  
\ee
\be
  e_{a} \oCt   A^a ,
  e^{a} \oCt  \A_a 
  \ee
\be
    e_{a} \Ct   (\y^a C)
,   e^{a} \oCt  (\oy_a  \oC)
  \ee
Here are the  mappings in $d_1$ that involve these terms:
    \be
    e_{a} \Ct   A^a  \stackrel{d_1= C \na}{\lra}   e_{a} \Ct   (\y^a C)
\ee
\be  e^{a} \oCt  \A_a   \stackrel{d_1= \oC \ov \na}{\lra}    e^{a} \oCt  (\oy_a  \oC)
  \ee
These four objects are `killed' in $E_1$, and so they do not survive to $E_2$.  
  We can regard this as being  the consequence of the \ED.    However there is no mapping in $d_1$ that includes the following two objects:
\be
    e^{a} \Ct   \A_a \in E_2 
 \ee
\be
  e_{a} \oCt   A^a  \in E_2 
   \ee
This lack of a mapping here arises from the fact that  $\Ct \oC$ and $\oCt C$ do not exist in the cohomology of $\d_{\rm Structure}$. t Since there is nothing else of this dimension with spin zero, the operators $d_r, r= 2, 3\cdots$ must all be zero in this sector.  For  example,  nothing in this sector can satisfy the relations:
 \be
 \lt [ N_{\rm Grading} , d_r\rt ] = r d_r, r= 2, 3\cdots
 \ee
 \be
 \lt [ N_{\rm ghost } , d_r\rt ] =  d_r, r=   2, 3\cdots
 \ee
 Hence it must be true that 
      \be
    e^{a} \Ct   \A_a \in E_{\infty} 
 \ee
     \be
    e_{a} \oCt   A^a \in E_{\infty} 
 \ee
   Now we can guess the isomorphism $ E_{\infty} \ra \cH$ for these two simple cases.  It is simply 

   \be
    e^{a} \Ct   \A_a \in E_{\infty}\lra \int d^4 x  e^{a} \oF_a\in \cH 
 \ee
\be
  e_{a} \oCt   A^a \in E_{\infty} \lra \int d^4 x  e_{a} F^a\in \cH 
 \ee 
 One can see by inspection that these are both cocycles and not coboundaries of the operators in equation (\ref{brstransWZ}).
 So the spectral sequence works for this simplest of examples.

\subsection{\Large The  Sector for dimension =-1 and spin =0}
\la{dimmin1nophi}
\refstepcounter{orange}
{ \theorange}.\;The next simplest example of $E_1$  with spin zero has dimension =-1.  Here we find  the following 14 terms with dimension =-1 and spin =0:

 \be
e_a (C \x \oC)  A^a ;
e^a (C \x \oC)  \A_a
\ee

  \be
 e^{ab} \Ct     \A_a \A_b 
; e^{a}_{b} \Ct     \A_a A^b 
; e_{ab} \Ct     A^a A^b 
\ee
  \be
 e^{ab} \oCt     \A_a \A_b 
; e^{a}_{b} \oCt     \A_a A^b 
; e_{ab} \oCt    A^a A^b 
\ee
The next examples are limited by the fact that  $\Ct \oC$ and $\oCt C$ do not exist in the cohomology of $\d_{\rm Structure}$:
\be
  e_a^{b}  \Ct    (C\y^a) \A_b ;   e_{ab}  \Ct    (C\y^a) A^b ;  e_{ab}  \Ct    (C\y^a)   (C\y^b)   
\ee
\be
  e^a_{b}  \oCt    (\oC\oy_a) A^b ;   e^{ab}  \oCt    (\oC\oy_a) \A_b ;  e^{ab}  \oCt    (\oC\oy_a)   (\oC\oy_b)   
\ee
 Here are the  mappings from $d_1$
\be
   e_{a}^{b}  \Ct    A^a  \A_b   \stackrel{d_1= C \na}{\lra} 
   e_{(a b)}  \Ct      (C\y^a)    \A_b  
 \ee
 \be
  e^{a}_{b} \oCt     \A_a A^b \stackrel{d_1=\oC \ov \na}{\lra}  
  e^{a}_{b} \oCt     (\oC\oy_a)   A^b
  \ee   
 \be
  e_{(a b)}  \Ct    A^a  A^b  \stackrel{d_1= C \na}{\lra} 
   e_{(a b)}  \Ct   A^a  (C\y^b)   
\ee
 \be
  e_{[a b]}  \Ct    (C\y^a) A^b  \stackrel{d_1= C \na}{\lra} 
   e_{[a b]}  \Ct    (C\y^a)   (C\y^b)   
\ee
  \be
    e^{(ab)} \oCt   \A_a   \A_b \stackrel{d_1=\oC \ov \na}{\lra}  
       e^{(ab)} \oCt   \A_a    (\oC\oy_b)  
\ee 
  \be
    e^{[ab]}\oCt   \A_a    (\oC\oy_b)    \stackrel{d_1=\oC \ov \na}{\lra}  
       e^{[ab]} \oCt    (\oC\oy_a)     (\oC\oy_b)  
\ee 
 \refstepcounter{orange}
{ \theorange}.\;  These  objects do not survive to $E_2$.  
   These mappings remove  the above objects, including both symmetries of the ones that have two similar indices.
  We can regard the above results as being examples of the \ED.

\refstepcounter{orange}
{ \theorange}.    However there is no mapping using $d_1$ that includes any of the following four  objects, so they survive to $E_2$:
\be
 e_a (C \x \oC)  A^a \in E_2
 ;
\ov e^a (C \x \oC)  \A_a \in E_2
\eb
 \ov e^{ab} \Ct     \A_a \A_b  \in E_2
 ;  e_{ab} \oCt    A^a A^b  \in E_2 
 \ee
 This is a consequence of the fact that the terms $(C \x \oC) C$ and $(C \x \oC) \oC$ and the terms $\Ct \oC$ and $\oCt C$  do not exist in the cohomology  of  $\d_{\rm Structure}$.

\refstepcounter{orange}
\la{d3paragraph}
{ \theorange}. But the story does not end here.  There are mappings that do map between these objects. There is no chance of making an operator $d_2$ here because
the difference in the value of $ N_{\rm Grading}$ is 3:
\be
 N_{\rm Grading} e_a (C \x \oC)  A^a = 5  e_a (C \x \oC)  A^a 
 \ee
\be
N_{\rm Grading}  \ov e^a (C \x \oC)  \A_a =5 e^a (C \x \oC)  \A_a 
\ee
\be
N_{\rm Grading}  e^{ab} \Ct     \A_a \A_b  =8 e^{ab} \Ct     \A_a \A_b 
\ee
\be
N_{\rm Grading}  e_{ab} \oCt    A^a A^b  = 8 e_{ab} \oCt    A^a A^b 
 \ee
 and this is not consistent with
 the equation
  \be
 \lt [ N_{\rm Grading} , d_r\rt ] = r d_r, r=0,1, 2, 3\cdots
 \ee
 for an operator $d_2$. It follows that these survive to the space $E_3$:
 \be
 e_a (C \x \oC)  A^a \in E_3
 ; 
 \ov  e^a (C \x \oC)  \A_a \in E_3
; \eb
  e^{ab} \Ct     \A_a \A_b  \in E_3
;  e_{ab} \oCt    A^a A^b  \in E_3
 \ee

However, we  can construct an operator $d_3$ as follows:
\[
d_3= \P_3 \d_1 \d_0^{\dag}\d_1 \d_0^{\dag} \d_1\P_3+*
\]\[ = \P_3  \og^{abc} \A_b \A_c  \ov\Lam^{a \dag}) (F^b \ov\Lam^{b \dag} )^{\dag}
 (F^d C \y^{d \dag})  ( C \oC \x^{\dag})^{\dag} (C\y_b) A^{b \dag}
\P_3+*\]\be= \P_3   \og^{abc} \A_b \A_c 
   A^{a \dag} (C \x)_{\dot \d}  \oC_{\dot \d}^{\dag}
\P_3+*
\la{dthree}
\ee

So then we get 
\be
e_a (C \x \oC)  A^a \stackrel{d_3}{\lra} e_a \og^{abc}  \Ct     \A_b \A_c 
\la{dimmonefirst}\ee
and
\be
\ov e^a (C \x \oC)  \A_a \stackrel{d_3}{\lra} e^a g_{abc}  \Ct     A^b A^c 
\la{dimmonesec}
\ee
and we also get the adjoint equations:
\be
e^{ab} \Ct     \A_a \A_b  \stackrel{d_3^{\dag}}{\lra} e^{ab} g_{abc}  (C \x \oC)  A^c
\la{dimmonefirstad}\ee
\be
e_{ab} \oCt     A^a A^b  \stackrel{d_3^{\dag}}{\lra} e_{ab} \og^{abc}  (C \x \oC)  \A_c
\la{dimmonesecad}
\ee
So the constraint equations here are:
\be 
e_a \og^{abc}=0
\ee
\be 
\ov e^a g_{abc}=0
\ee
and for the adjoint equations, the constraint equations are
\be
e^{ab} g_{abc}  =0
\ee
\be
e_{ab} \og^{abc}=0.
\ee

\refstepcounter{orange}
{ \theorange}. What do these equations mean? These are constraint equations. Any solutions of these equations are elements of the space $E_4= \ker d_3\cap \ker d_3^{\dag}\cap E_3$.  Since there are no further $d_r$ here, they are also elements of $E_{\infty}$.  They have various solutions, depending on the particular form of the tensors involved.  Thus, for example, if there is only one kind of field so that a=1 only, then these equations are quite simple. But if there are two (or more) kinds of field so that $a=1,2$ (or $a=1\cdots n$), then there are plenty of different solutions. There are no more differentials $d_r,r\geq 4$
 in this sector because   all the difference in grading among the surviving terms is used up by this operator $d_3$. 
But sometimes, when the constraints are satisfied,  there are objects that survive into $E_4= E_{\infty}$ and for these we get the isomorphisms
\be
e_a (C \x \oC)  A^a \in E_{\infty}\ra  \int d^4 x  e_{a} \ov \G^a \in \cH 
\la{dimmonefirstch}
\ee
\be
e^a (C \x \oC)  \A_a \in E_{\infty}\ra  \int d^4 x  e^{a}  \G_a \in \cH 
\ee
Now these are in $E_{\infty}\ra   \in \cH $ only when the constraint equations are true. For example, this is easy to grasp, for the first ones, by looking at the transformations in (\ref{brstransWZ}), which include:
\be
\d \G_i 
\eb
=
  \Box    {\ov  A}_{i} 
  +   m {g}_{iq} F^q  +2 g_{ijk} ({ A}^{j}{ F}^{k}- {\y}^{j\a}{\y}_{\a}^{k})
 -\pa_{ \a \dot \b } Y_{i}^{ \a}    {\ov C}^{\dot \b}   
+ \x^{\g \dot \d} \partial_{\g \dot \d} \G_i
\ee
because we can see that
\be
\d
\int d^4 x e^i \G_i 
\eb=
\int d^4 x e^i \lt \{  \Box    {\ov  A}_{i} 
  +   m {g}_{iq} F^q  +2 g_{ijk} ({ A}^{j}{ F}^{k}- {\y}^{j\a}{\y}_{\a}^{k})
\ebp-\pa_{ \a \dot \b } Y_{i}^{ \a}    {\ov C}^{\dot \b}   
+ \x^{\g \dot \d} \partial_{\g \dot \d} \G_i
\rt \}
\ee
and for the massless case this is
\be
\d
\int d^4 x e^i \G_i 
=
\int d^4 x e^i \lt \{  
  2 g_{ijk} ({ A}^{j}{ F}^{k}- {\y}^{j\a}{\y}_{\a}^{k})
\rt \} =0 
\ee
if 
\be
\ov  e^i 
 g_{ijk}
=0
\ee

\refstepcounter{orange}
{ \theorange}.\;Note that these involve the pseudofields $\G_a,\ov \G^a$, even though these were eliminated from the spectral sequence. This was mentioned above.  It is a very important feature of the BRS cohomology of this theory, and it reappears in sections \ref{phiintro}, 
 \ref{dimzerosec} and 
\ref{nightmare} in a very important way. 
The other isomorphisms here are:
\be
e^{(ab)} \Ct     \A_a \A_b  \in E_{\infty}\ra  \int d^4 x e^{(ab)} 
\lt (  \A_a \oF_b - \fr{1}{2}\oy^{\dot \a}_a \oy_{b\dot \a} \rt ) 
\in \cH 
\la{dimmonefirstadch}
\ee
\be
e_{(ab)} \oCt     A^a A^b  \in E_{\infty}\ra  \int d^4 x e_{(ab)} 
\lt ( \A^a F^b - \fr{1}{2}\y^{a \a} \y^{b}_{ \a} \rt ) 
\in \cH 
\ee
 These are for tensors that satisfy the constraint equations, of course.

\subsection{\Large $E_1$  in the  Sector for dimension = 0 and spin = 0}

\refstepcounter{orange}
{ \theorange}.\;From the previous two sections we can see that things are getting more complex as the dimension increases.  Things also get more complicated as the spin increases, as we shall start to see in section \ref{phiintro} below. 

Now at dimension zero we get a new kind of object, namely the following:
\be
(C \x \oy_a) \eb
(C \x)^{\dot \a}  \A_{a\dot \a \b} C^{\b}
\ee
\be
(C \x)^{\dot \b}  A_{a \b \dot \b} C^{\b}\eb
(C \x)^{\dot \a}  \y^{a }_{(\a \b) \dot \a} C^{\a} C^{\b},
\ee
and the following
\be
(\oC \x \y^a) \ee
\be
(\oC \x)^{ \b}  \A_{a \b \dot \b} \oC^{\dot \b}\ee
\be
(\oC \x)^{ \a}  A^a_{\a \dot \b} \oC^{\dot\b}
\ee
\be
(\oC \x)^{ \b}  
\oy_{a  (\dot \a\dot \b), \b} \oC^{\dot \b}\oC^{\dot \a},
\ee
The following maps come from some of the first terms in $d_1$:
\be
(C \x \oy_a) \stackrel{d_1}{\lra}
(C \x)^{\dot \a}  \A_{a\dot \a \b} C^{\b}
\ee
\be
(\oC \x \y^a) \stackrel{d_1}{\lra}
(\oC \x)^{ \a}  A^a_{\a \dot \b} \oC^{\dot\b}
\ee
\be
(C \x)^{\dot \b}  A^a_{ \b \dot \b} C^{\b}\stackrel{d_1}{\lra}
(C \x)^{\dot \a}  \y^{a }_{\a, \b \dot \a} C^{\a} C^{\b},
\ee
\be
(C \x)^{\dot \b}  \A_{a \b \dot \b} C^{\b}\stackrel{d_1}{\lra}(\oC \x)^{ \b}  \oy_{a \dot \a, \b \dot \b} \oC^{\dot \b}\oC^{\dot \a},
\ee
So these eight terms are gone from $E_2$.
We also have:
\be
(C \x \oC) AA ,\eb
(C \x \oC) \A\A ,\eb
(C \x \oC) A \A 
\ee 
Here we are getting lazy and we drop some indices.  They will be replaced when we need them.
These terms are not mapped by $d_1$. So they survive to $E_2$.

\subsection{\Large Use of the operator Approach}

\refstepcounter{orange}
{ \theorange}.\;Rather than write all the possibilities for the following terms, we will now adopt the form that conforms to our results for $E_2$ in section 
\ref{operatorchapter}, which follows: 
\be
 e^{abc} \Ct    \A_a \A_b \A_c ,\eb
 e_{ab} \Ct    \y^a \y^b\ee

 and
 \be
  e_{abc}  \oCt A^a A^b A^c ,\eb
   e^{ab} \oCt    \oy_a \oy_b
 \ee 

Now we do get a map from $d_2$ here\footnote{This map comes from $d_2 = \P_2\d_1 d_0^{\dag} \d_1 \P_2=\P_2( C \y A^{\dag}) 
(C \oC \x^{\dag})^{\dag} ( C \y A^{\dag})  \P_2$ }
\be
 e_{ab}  (C \x \oC) A^aA^b  \stackrel{d_2}{\lra}
 e_{ab} \Ct    (\y^a \y^b)
\la{weirdd2}\ee
Similarly
\be
(C \x \oC) e^{ab} \A_a\A_b\stackrel{d_3}{\lra}e^{ab} \oCt    (\oy_a \oy_b)
\ee

\refstepcounter{orange}
{ \theorange}.\;
Next we note that the same issue arises again for the terms where  we got  a mapping with $d_3$ that arose for the analogous terms in dimension -1 in
(\ref{dimmonefirst}) and (\ref{dimmonesec}) using the formula in 
(\ref{dthree}):
\be
e_f^d (C \x \oC)  A^f \A_d\stackrel{d_3}{\lra} e_f^d \og^{fbc}  \Ct     \A_b \A_c \A_d
+e_f^d g_{dbc}  \oCt   A^f  A^b A^c 
\ee
and we also get the adjoint equations analogous to (\ref{dimmonefirstad}) and (\ref{dimmonesecad}): 
\be
e_f^d (C \x \oC)  A^f \A_d \stackrel{d_3^{\dag}}{\longleftarrow} 
 \ov e^{bcd} g_{bcf}  \Ct     \A_b \A_c \A_d
+\og^{bcd} e_{bcf}    \oCt   A^f  A^b A^c 
\ee
As in the case of dimension -1, these are constraint equations with many solutions. 

\refstepcounter{orange}
{ \theorange}.\;When the constraint equations are solved, there are objects that survive into $E_{\infty}$ and for these we get the isomorphisms
\be
e_a^b (C \x \oC)  A^a \A_b \in E_{\infty}\ra
\eb
 e_a^b \int d^4 x \lt \{  \lt ( \ov \G^a  \A_b+ \ov Y^{a\dot \b} \oy_{b \dot \b} + \ov \Lam^a \ov F_b  \rt )
 \ebp
 -    \lt ( \G_b  A^a+ Y^{\b}_b \y^a_{ \b} +  \Lam_b F^a  \rt ) 
\rt \}\in \cH  \ee
\be 
\ov e^{(abc)} \Ct     \A_a \A_b\A_c  \in E_{\infty}\ra  \int d^4 x \ov e^{(abc)} 
\lt (  \A_a  \A_b \oF_c -  \A_a \oy^{\dot \a}_b \oy_{c\dot \a} \rt ) 
\in \cH 
\ee
\be
e_{(abc)} \oCt     A^a A^b A^c  \in E_{\infty}\ra  \int d^4 x e_{(abc)} 
\lt ( A^a  \A^b F^c -A^a \y^{b \a} \y^{c}_{ \a} \rt ) 
\in \cH 
\ee
 These are for tensors that satisfy the constraint equations, of course.

\section{\LARGE An operator approach to part of the Space $E_1$ and the operator $d_1=\P_1 \d_1 \P_1$ and the space $E_2= \ker \D_1$ }
\la{operatorchapter}

\refstepcounter{orange}
{ \theorange}.\;The \ED\ becomes very complicated as the dimension and spin increase.  We can tame it, partly, with the following operator methods.
We recall from   the discussion of the space $E_1$ in section \ref{Spssection} that:
\ben
\item
The only fields or pseudofields that appear in $E_r, r \geq 1$ are the totally symmetrized 
fields
\be
A^a, A^a_{\a_1,\dot \a_1}, \cdots A^a_{(\a_1\cdots \a_n ),(\dot \a_1\cdots \dot \a_n )}, n=2,3,\cdots
\ee
\be
\y_{\a}^a, \y^a_{(\a \a_1),\dot \a_1}, \cdots \y^a_{(\a \a_1\cdots \a_n ),(\dot \a_1\cdots \dot \a_n )}, n=2,3,\cdots
\ee
and their complex conjugates
\be
\A_a, \A_{a, \a_1,\dot \a_1}, \cdots \A_{a,(\a_1\cdots \a_n ),(\dot \a_1\cdots \dot \a_n )}, n=2,3,\cdots
\ee
\be
\oy_{a, \dot \a} , \oy_{a, (\dot \a \dot \a_1), a_1}, \cdots \oy_{a,(\dot \a \dot \a_1\cdots \dot  \a_n ),( \a_1\cdots   \a_n )}, n=2,3,\cdots
\ee

\item
The spacetime constant ghosts $C_{\a},\oC_{\dot \a}, \x_{\a\dot \b}$ appear only in the form dictated by the cohomology of the operator 
\be
\d_{\rm Structure} = C_{\a}\oC_{\dot \a} \x_{\a\dot \b}^{\dag}
\ee
\een

\subsection{\Large The operator $d_1$ for the \wzcss\ theory}
\la{nanoprime}
\refstepcounter{orange}
{ \theorange}.\;Because the subspace $E_1$ is limited to a dependence on just the symmetrized fields $A,\y$, and their derivatives, and complex conjugates, with no auxiliaries and no pseudofields, we find that only a small piece of the operators $\na^{\rm Big}$ in (\ref{naprime}) and (\ref{naprimebar}), and nothing from the operator (\ref{eqmoterms}), are relevant for $d_1$, and we will denote the relevant pieces by simply $\na$ without the word Big on it: 

\be 
d_1 = C^{\a} \na_{\a}+ \oC^{\dot \a} \ov \na_{\dot\a}
\ee
where
\[
 C^{\a} \na_{\a}
= C^{\a} \P_1 \lt \{
 \y^a_{\a} A^{a \dag} 
 +\y^a_{(\a \b) \dot \b} A_{\b \dot \b}^{a \dag} 
 + \y^a_{(\a \b \g), (\dot \b\dot \g )} A_{ (\b \g) , (\dot \b\dot \g )}^{a \dag} + \rt.\cdots\]
\be\lt. +  
\A_{a \dot \b \a }  \oy^{\dag}_{a \dot \b }
+  
\A_{a (\dot\a\dot \b), (\a \b) }  \oy^{\dag}_{a (\dot \a \dot \b),\b   }
+ \cdots
\rt \} \P_1
\la{formofna}
\ee 
and the adjoint is 
\be 
 C^{\m \dag} \na_{\m}^{\dag} = \P_1
 C^{\m \dag}\lt \{
 \y^b_{\m} A^{b \dag} 
 +\y^a_{(\m \n) \dot \n} A_{\n \dot \n}^{b \dag} 
 + \y^b_{(\m \n \lam), (\dot \n\dot \lam )} A_{ (\n \lam) , (\dot \n\dot \lam )}^{b \dag} + \cdots 
  \ebp
+  
\A_{b \dot \n \m } \oy^{\dag}_{b \dot \n }
+  
\A_{b(\dot\m\dot \n), (\m \n) }  \oy^{b\dag}_{b (\dot \m \dot \n),\n   }
+ \cdots
\rt \}^{\dag}\P_1
\ee
We use different indices on these because they are going to be put together.

\subsection{\Large Mappings in $E_1$}
\refstepcounter{orange}
{ \theorange}.\;If we look at the mappings by $d_1$ in detail we see that, (mostly) abstracting from the presence of the ghosts, and any contractions, and ignoring the presence of the projection operators $\P_1$, we have mappings that look like this:
\be
M_0 \stackrel{  \na_{\a} }{\lra }  P_{1 }  
\ee
\be
M_0 \stackrel{  \ov\na_{\dot \a} }{\lra }  \oP_{1} 
\ee

\be
 P_{n}  \stackrel{  \na_{\a} }{\lra } P_{n+1} , n = 1\cdots \infty
\ee
\be
\oP_{n}  \stackrel{  \ov\na_{\dot \a} }{\lra }   \oP_{n+1}, n = 1\cdots \infty 
\ee

\be
 Q_{0}= (\x \oC)_{ \a}  Q^{0, \a}  \stackrel{\oC^{\dot \a} \ov\na_{\dot \a} }{\lra }    ( C \x \oC) R_0  
\ee
\be
 \oQ_{0}= (\x \oC)_{ \a} \oQ^{0, \a}  \stackrel{C^{\a} \na_{\a} }{\lra }  ( C \x \oC) R_0  
\ee

\be
 R_0  \stackrel{d_1 }{\lra }    0
 \ee

\be
 Q_{n}  \stackrel{  \na_{\a} }{\lra }   Q_{n+1}, n=0,1,2,\cdots \infty
\ee 
\be
 \oQ_{n}   \stackrel{ \ov\na_{\dot \a} }{\lra }  \oQ_{n+1}, n=0,1,2,\cdots \infty
\ee

\be
 R_{n}  \stackrel{  \na_{\a} }{\lra }   R_{n+1}, n=0,1,2,\cdots \infty
\ee 
\be
 \oR_{n}   \stackrel{  \ov\na_{\dot \a} }{\lra }  \oR_{n+1}, n=0,1,2,\cdots \infty
\ee

\refstepcounter{orange}
{ \theorange}.\;
There are troublesome bits around $M_0, Q_0,\ov Q_0, R_0$, because both parts of $d_1=\P_1\lt \{ C^{\a} \na_{\a}+\oC^{\dot \a} \ov\na_{\dot \a}\rt \} \P_1$ get involved in the mappings there.  We will continue to use the \ED\ for those parts.  However one location which is not troubled is the action of $d_1$ on the $ \Ct$ and the $ \oCt$ sectors. There is no mixing of $C^{\a} \na_{\a}$  and $\oC^{\dot \a} \ov\na_{\dot \a}$ for either of those.  The same is true for some of the other sectors where only C is involved.  We will denote those sectors by $E_{2,C} \subset E_{1,C} $.  In this paper we will only use these results for the $\Ct$ sector, because the other parts are not very complicated anyway,  at these low values of the dimension.  At higher levels of dimension we would want to reconsider this. 

\refstepcounter{orange}
{ \theorange}.\;
So, as a result, for the special sector $ \Ct$ where $d_1 = C^{\a} \na_{\a}$,  we can evaluate the Laplacian operator $\D_1=\lt ( d_1 + d_1^{\dag}\rt )^2$ for just the C part of $d_1$.  Note the initial simplification which follows from $d_1^2=0$.  We use indices from the middle of the alphabet so that they do not get mixed, by repetition, with indices from the beginning of the alphabet.  We willl drop the projection operators $\P_1,\P_2$ for the time being, but we return to them when we need them. 
\be
\D_1=\lt ( d_1 + d_1^{\dag}\rt )^2
=
\lt \{ C^{\a} \na_{\a}, C^{\m \dag} \na_{\m}^{\dag}\rt \} 
\eb=
  C^{\a} \na_{\a} C^{\m \dag} \na_{\m}^{\dag} 
+
 C^{\m \dag} \na_{\m}^{\dag}  C^{\a} \na_{\a}
\eb
=
\d^{\a}_{\m}   \na_{\m}^{\dag}  \na_{\a}
+
   C^{\a}  C^{\m \dag}  \lt \{ \na_{\a},\na_{\m}^{\dag}\rt \} 
\eb
= 
   \na_{\a}^{\dag}  \na_{\a}
\eb
+ \P_1
C^{\a} C^{\m \dag}
\lt \{
 \y^a_{\a} A^{a \dag} 
 +\y^a_{(\a \b) \dot \b} A_{\b \dot \b}^{a \dag} 
 + \y^a_{(\a \b \g), (\dot \b\dot \g )} A_{ (\b \g) , (\dot \b\dot \g )}^{a \dag} + \cdots 
\ebp
 +  
\A_{a\dot \b \a } \oy^{\dag}_{a\dot \b }
+  
\A_{a(\dot\a\dot \b), (\a \b) } \oy^{\dag}_{a (\dot \a \dot \b),\b   }
+ \cdots
  ,
\ebp
  A^{b }  \y^{b\dag}_{\m}
 + A_{\n \dot \n}^{b } \y_{(\m \n) \dot \n}^{b \dag}
 + A_{ (\n \lam) , (\dot \n\dot \lam )}^{b } \y_{(\m \n \lam), (\dot \n\dot \lam )}^{b\dag}+ \cdots 
\ebp +  
\oy_{b\dot \n }\A_{b\dot \n \m }^{ \dag} 
+  
\oy_{b (\dot \m \dot \n),\n   }\A_{b(\dot\m\dot \n), (\m \n) }^{ \dag} 
+ \cdots
 \rt \}
\ee 
\refstepcounter{orange}
{ \theorange}.\;Note the convenient way that this splits into small pieces as follows:
\be
= 
   \na_{\a}^{\dag}  \na_{\a}
\eb
+ \P_1
C^{\a} C^{\m \dag}
\lt \{
 \y^a_{\a} A^{a \dag} , A^{b }  \y^{b\dag}_{\m}\rt \}\eb
 + \P_1
C^{\a} C^{\m \dag}
\lt \{
\y^a_{(\a \b) \dot \b} A_{\b \dot \b}^{a \dag} , A_{\n \dot \n}^{b } \y_{(\m \n) \dot \n}^{b \dag}\rt \}
\eb
+ \P_1
C^{\a} C^{\m \dag}
\lt \{  \y^a_{(\a \b \g), (\dot \b\dot \g )} A_{ (\b \g) , (\dot \b\dot \g )}^{a \dag},A_{ (\n \lam) , (\dot \n\dot \lam )}^{b } \y_{(\m \n \lam), (\dot \n\dot \lam )}^{b\dag}\rt \}+\cdots
\eb
+ \P_1
C^{\a} C^{\m \dag}
\lt \{ \A_{\dot \b \a }^{a } \oy^{a\dag}_{\dot \b }
,\oy^{b}_{\dot \n }\A_{\dot \n \m }^{b \dag} 
\rt \}
\eb
+ \P_1
C^{\a} C^{\m \dag}
\lt \{ \A_{a(\dot\a\dot \b), (\a \b) } \oy^{\dag}_{a (\dot \a \dot \b),\b   }
,\oy_{a (\dot \m \dot \n),\n   }\A_{b (\dot\m\dot \n), (\m \n) }^{ \dag} 
\rt \}
\eb
+ \P_1
C^{\a} C^{\m \dag}
\lt \{ 
\cdots
\rt \}
\ee 
Now we can evaluate the terms one by one.
Here they are:

\subsection{\Large Zeroth Term:}
\be 
   \na_{\a}^{\dag}  \na_{\a}
\la{zeroth}
\ee

\subsection{\Large  $\y A $ Terms}
\refstepcounter{orange}
{ \theorange}.\;{The first $\y A $Term is:}

\be
 \P_1
C^{\a} C^{\m \dag}
\lt \{
 \y^a_{\a} A^{a \dag} , A^{b }  \y^{b\dag}_{\m}\rt \}
 \ee
\be
= \P_1
C^{\a} C^{\m \dag}
\lt \{
 \y^a_{\a} A^{a \dag}  A^{b }  \y^{b\dag}_{\m}+ 
 A^{b }  \y^{b\dag}_{\m} \y^a_{\a} A^{a \dag}  \rt \}
 \ee
\be
= \P_1
C^{\a} C^{\m \dag}
\lt \{
 \y^a_{\a}   \d^{b }_{a}  \y^{b\dag}_{\m}+ 
 A^{b }    \d^a_b \d_{\a}^{\m}  A^{a \dag}  \rt \}
 \ee
\be
= \P_1
\lt \{
(C^{\a} \y^a_{\a} )  (  C^{\m  }\y^{a}_{\m})^{\dag}   + 
 C^{\a}C^{\a \dag}   A^{a }     A^{a \dag}  \rt \}
 \ee
and this is the sum of two positive terms. 
 {The  second  $\y A $ Term  is}
 \be
 \P_1
C^{\a} C^{\m \dag}
\lt \{
\y^a_{(\a \b) \dot \b} A_{\b \dot \b}^{a \dag} , A_{\n \dot \n}^{b } \y_{(\m \n) \dot \n}^{b \dag}\rt \}
\ee
 and this is
 \be
=
 \P_1
\lt \{
C^{\a} C^{\m \dag}
\y^a_{(\a \b) \dot \b} 
   \y_{(\m \b) \dot \b}^{a \dag}
\ebp+C^{\a} C^{\m \dag}
A_{\n \dot \n}^{a }
 \fr{1}{2}\lt (  \d_{\a}^{\m} \d_{\b}^{ \n  } +\d_{\a}^{\n} \d_{\b}^{ \m}  \rt)
  \d_{  \dot \b}^{ \dot \n} 
 A_{\b \dot \b}^{a \dag}\rt \}
 \ee
 which is the sum of three postive terms
 \be
=
 \P_1
\lt \{
(C^{\a} 
\y^a_{(\a \b) \dot \b} )(
  C^{\m \dag} \y_{(\m \b) \dot \b}^{a \dag})
\ebp+ \fr{1}{2}( C^{\m} 
A_{\b \dot \b}^{a } )(C^{\m \dag}A_{\b \dot \b}^{a \dag} )    + \fr{1}{2} (C^{\a} 
A_{\a \dot \b}^{a })( C^{\m \dag} A_{\m \dot \b}^{a \dag})   
 \rt \}
 \ee
{The third $\y A $ Term is}
\be
 \P_1
C^{\a} C^{\m \dag}
\lt \{  \y^a_{(\a \b \g), (\dot \b\dot \g )} A_{ (\b \g) , (\dot \b\dot \g )}^{a \dag},A_{ (\n \lam) , (\dot \n\dot \lam )}^{b } \y_{(\m \n \lam), (\dot \n\dot \lam )}^{b\dag}\rt \}+\cdots
\ee
which is
the sum of three positive terms:
\be
= \P_1
C^{\a} C^{\m \dag}
\lt \{  \y^a_{ (\a \b \g), (\dot \b\dot \g )}
 \y_{(\m \b \g) , (\dot \b\dot \g )}^{a\dag}\rt \} 
\ee
\be
+
 \P_1
\lt \{ A_{( \b \g),  (\dot \b\dot \g )}^{a } 
\fr{1}{3} C^{\a} C^{\a \dag} 
  A_{ (\b \g) , (\dot \b\dot \g )}^{a \dag}
\rt \} 
\eb
+
 \P_1
\lt \{ A_{( \a \g),  (\dot \b\dot \g )}^{a } 
 \fr{2}{3}  C^{\a} C^{\b \dag}
 A_{ (\b \g) , (\dot \b\dot \g )}^{a \dag}
\rt \} 
\ee
\subsection{\Large  $\oy \A $ Terms}
\refstepcounter{orange}
{ \theorange}.\;{The 
First $\oy \A$ term  is}
\be
 \P_1
C^{\a} C^{\m \dag}
\lt \{ \A_{a\dot \b \a } \oy^{\dag}_{a\dot \b }
,\oy_{b\dot \n }\A_{b\dot \n \m }^{ \dag} 
\rt \}
\ee
which is
\be
= \P_1
\lt \{ 
\A_{aq\dot \b \a }
C^{\a} C^{\m \dag}
\A_{a\dot \b \m }^{\dag} 
+
\oy_{a\dot \n }
C^{\a} C^{\a \dag}
\oy^{\dag}_{a\dot \n }
\rt \}
\ee
which is two positive terms. {Then the second  $\oy \A $  Term is:}
\be \P_1
C^{\a} C^{\m \dag}
\lt \{ \A_{a(\dot\a\dot \b), (\a \b) }\oy^{\dag}_{a (\dot \a \dot \b),\b   }
,\oy_{ b(\dot \m \dot \n),\n   }\A_{b(\dot\m\dot \n), (\m \n) }^{ \dag} 
\rt \}
\ee
which is the sum of three postive terms:
\be \P_1
C^{\a}  \A_{a(\dot\a\dot \b), (\a \b) } 
\A_{a(\dot \a \dot \b), (\m \b) }^{ \dag} C^{\m \dag}
\eb
+\P_1
\fr{1}{2}
 C^{\a}  C^{\a \dag}
\oy_{a (\dot\a\dot \b),\b   }    \oy^{\dag}_{ a(\dot \a \dot \b),\b   }
\eb
 +\P_1
\fr{1}{2} C^{\n}  \oy_{a (\dot\a\dot \b),\n   }  
C^{\b \dag}
  \oy^{\dag}_{a (\dot \a \dot \b),\b   } 
\ee
\subsection{\Large Collect the dominant equations}
\refstepcounter{orange}
{ \theorange}.\;We can summarize the foregoing results by collecting the equations that imply the other equations:
\be
(C^{\a} \y^a_{\a} )  (  C^{\m  }\y^{a}_{\m})^{\dag}  E_{2,C}=0
 \ee
\be
  C^{\a}C^{\a \dag}   A^{a }     A^{a \dag}   E_{2,C}=0
 \ee
 \be
C^{\a} 
\y^a_{(\a \b) \dot \b} 
  C^{\m \dag} \y_{(\m \b) \dot \b}^{a \dag} E_{2,C}=0
  \ee
\be
  C^{\m} 
A_{\b \dot \b}^{a } C^{\m \dag}A_{\b \dot \b}^{a \dag}    
E_{2,C}=0
\ee
\be
C^{\a} C^{\m \dag}
  \y^a_{ (\a \b \g), (\dot \b\dot \g )}
 \y_{(\m \b \g) , (\dot \b\dot \g )}^{a\dag}E_{2,C}=0
\ee
\be
A_{a( \b \g),  (\dot \b\dot \g )} 
  C^{\a} C^{\a \dag} 
  A_{a (\b \g) , (\dot \b\dot \g )}^{\dag}E_{2,C}=0
  \ee
\be
\A_{a\dot \b \a }
C^{\a} C^{\m \dag}
\A_{a\dot \b \m }^{ \dag} 
E_{2,C}=0
\ee
\be
\oy_{a\dot \n }
C^{\a} C^{\a \dag}
\oy^{\dag}_{a\dot \n }
E_{2,C}=0\ee
\be  
C^{\a}  \A_{a(\dot\a\dot \b), (\a \b) } 
\A_{a(\dot \a \dot \b), (\m \b) }^{ \dag} C^{\m \dag}
E_{2,C}=0
\ee
\be
 C^{\a}  C^{\a \dag}
\oy_{ a(\dot\a\dot \b),\b   }    \oy^{\dag}_{a (\dot \a \dot \b),\b   }
E_{2,C}=0
\ee
\subsection{\Large The General Terms}
\refstepcounter{orange}
{ \theorange}.\;One can easily continue this series.  The general $\y A$ term yields
\be
 (C \x^2 C)  \P_2  (C^{\m \dag}
   \y_{(\m \b_1 \cdots \b_n ) , (\dot \b_1 \cdots \dot \b_n )}^{a\dag} )E_{2,C} =0; n = 0,1,2\cdots 
\ee
\be
(C \x^2 C)  \P_2 
N_C  N_A E_{2,C} =0
  \ee 
  The general $\oy \A$ term yields
\be  
(C \x^2 C)  \P_2  \A_{a(\dot \a_1 \cdots\dot \a_n), (\m \b_2 \cdots \b_n  )}^{\dag} C^{\m \dag}
E_{2,C}=0; n = 1,2\cdots 
\ee
 \be
(C \x^2 C)  \P_2 
N_C  N_{\oy} E_{2,C} =0
\ee
\subsection{\Large Summary of the equations}

\la{fromdonefore2}

\refstepcounter{orange}
{ \theorange}.\;Let us start with the first term above
\be
= \P_1
\lt \{
(C^{\a} \y^a_{\a} )  (  C^{\m  }\y^{a}_{\m})^{\dag}   + 
 C^{\a}C^{\a \dag}   A^{a }     A^{a \dag}  \rt \}
 \ee
and the fourth term above (because it is also simple)
\be
= \P_1
\lt \{ 
\A_{a\dot \b \a }
C^{\a} C^{\m \dag}
\A_{a\dot \b \m }^{ \dag} 
+
\oy_{a\dot \n }
C^{\a} C^{\a \dag}
\oy^{\dag}_{a\dot \n }
\rt \}
\ee 
 The sum is postive so we know in particular that these mean that

\be
 C^{\a}C^{\a \dag}   A^{a }     A^{a \dag}  E_{2,C}=0
\la{first} \ee
\be
\oy_{a\dot \n }
C^{\a} C^{\a \dag}
\oy^{\dag}_{a\dot \n }
E_{2,C}=0
\la{second}\ee
and
\be
(C^{\a} \y^a_{\a} )  (  C^{\m  }\y^{a}_{\m})^{\dag} E_{2,C}=0
\la{third}
\ee
and
\be 
\A_{a\dot \b \a }
C^{\a} C^{\m \dag}\A_{a\dot \b \m }^{ \dag} E_{2,C}=0
\la{fourth}\ee

\subsection{\Large Meaning of the equations in section \ref{fromdonefore2}}
\la{explanationofdoneande2}

\refstepcounter{orange}
{ \theorange}.\; The first term $\na_{\a} E_{2,C}=0$ from (\ref{zeroth}) 
  is quite easy to satisfy and we do not need to worry about it in the following. 
For example, looking at (\ref{formofna}), $\na_{\a} E_{2,C}=0$ is true for any terms that do not contain $A, \oy$ or their derivatives.

\refstepcounter{orange}
{ \theorange}.\;The equations (\ref{first}) and (\ref{second}), and their generalizations,  
are easy to understand. The equation (\ref{first})   means
that if $E_{2,C}$ contains $A^{a }$   then it does not contain $C$.
Note that there is no equation at all that combines the underived field  $\A$ and C.
So we can get arbitrary powers of the underived field $\A_a$ even if C is present.  Note that here we mean C that is not contained in $\Ct$.  
The reason for that is that there is a projection operator $\P_1$ here that ensures that this contraction remains in place and is not counted as a $C$ term for the purposes of this reasoning. 

\refstepcounter{orange}
{ \theorange}.\;
Clearly we see that if C is present then $A, A_{\a \dot \b} A_{\a\b \dot \a \dot\b}$ are not present and $\oy_{\dot \b} , \oy_{ (\dot \b\dot \d), \a} \oy_{(\dot \a \dot\b\dot\g), (\a\b) }$ are not present.
The  equations (\ref{third}) and (\ref{fourth}),  and their generalizations,   are a little more obscure. But they are also easy to understand.   The  equation (\ref{third}) means that if   $E_{2,C}$ contains $\y ^{a}_{\m}$   and also $C$, then they appear in a combination such that the contraction yields zero.  But that simply means that the spin is maximal, which means they appear in a form where all the undotted indices on these two fields are symmetrized. The simplest example of this is
\be
E_{\rm 2\; example} = \A_{\dot \b \b} C_{\a}  + \A_{\dot \b \a} C_{\b} 
\ee
Then we note that
\be
\A_{\dot \d \g}^{\dag} C_{\d}^{ \dag}\ve_{\g \d}
\ee
 yields zero on this symmetrized form.
In fact (dropping the index for now)
\be
\A_{\dot \d \g}^{\dag} C_{\d}^{ \dag} \ve_{\g \d}E_{\rm 2\; example} \eb
=\A_{\dot \d \g}^{\dag} C_{\d}^{ \dag}\ve_{\g \d}
\lt \{ \A_{\dot \b \b} C_{\a}  + \A_{\dot \b \a} C_{\b} 
\rt \}
\ee

\be
=\A_{\dot \d \g}^{\dag} C_{\d}^{ \dag}\ve_{\g \d}
\lt \{ \A_{\dot \b \b} C_{\a}  + \A_{\dot \b \a} C_{\b} 
\rt \}\ee
\be
=\ve_{\g \d}
\lt \{ \A_{\dot \d \g}^{\dag} C_{\d}^{ \dag}\A_{\dot \b \b} C_{\a}  + \A_{\dot \d \g}^{\dag} C_{\d}^{ \dag}\A_{\dot \b \a} C_{\b} 
\rt \}\ee
\be
=\ve_{\g \d}
\lt \{ \d^{\dot \d \g}_{\dot \b \b} \d^{\d}_{\a}  +
 \d^{\dot \d \g}_{\dot \b \a} C^{\d}_{\b} 
\rt \}\ee
\be
= \d^{\dot \d}_{\dot \b}  \ve_{\g \d}
\lt \{\d^{ \g}_{ \b} \d^{\d}_{\a} +  \d^{ \g}_{ \a} \d^{\d}_{\b}   
\rt \}\ee
\be
= \d^{\dot \d}_{\dot \b}\lt \{ \ve_{\b \a}+ \ve_{\a \b} 
\rt \}
=0
\ee
So indeed these equations all imply maximum spin just as in \ci{dixminram}.

\subsection{\Large Higher Values of   dimension   and spin for the expressions}

\refstepcounter{orange}
{ \theorange}.\;We know that there are objects with symmetrized spin if they contain the expression $\Ct$ and also the fields $\y$ or $\A$ and $C$, except that this does not apply to $\A$ with no derivatives.  
 Do we know all of the $d_r$ for these?  We can figure the $d_r$ out for any given subspace of low dimension.  But perhaps things become more complicated as the dimension increases?  If these are relevant to the superstring, maybe they have interesting cohomology?  

\refstepcounter{orange}
{ \theorange}.\;
  In the next section, we will  introduce a constant spinor so that we can look at the parts of that BRS cohomology with  spin $\fr{1}{2}$, by restricting ourselves to Lorentz invariant objects that include one of these constant spinors.    
In this paper we will not go beyond the spin $\fr{1}{2}$ objects. 
Nor will we try to probe the meaning of the various constraint equations.
Those questions are for a future effort.  

\refstepcounter{orange}
{ \theorange}.\;
However now we have the full problem including the pseudofields, and they make an important difference.  
Whereas in \ci{dixminram} we could see only the possible anomalies, now we can also see the operators that can be anomalous.  The reason they were not obvious in \ci{dixminram} is that they depend crucially on the pseudofields--indeed the pseudofields are needed to construct these new invariants, as we will see in  section \ref{phiintro}. But all of them have non-zero spin.  
So now, in section \ref{phiintro}, we will introduce objects that can be contracted with the spin to form Lorentz invariants and anomalies that are spinless. 

\section{\LARGE The part of the cohomology that has  spin $ \fr{1}{2}$:
Addition of a  constant spinor $\f_{\dot \a}, \ov \f_{\a}$ with 
  $\rm dimension =  \fr{1}{2}$}
  \la{phiintro}
\refstepcounter{orange}
{ \theorange}.\;  We know that there is cohomology with a free spinor index.  So to probe this, we will contract the free spinor index with a constant spinor $\f_{\dot \a} $ (or its complex conjugate $ \ov \f_{\a}$). We will attribute a dimension 
 $\rm dimension =  \fr{1}{2}$ to it. We look at terms that have either  one constant spinor  $\f$ or one constant spinor  $ \ov \f$. We exclude terms that consist only of   ghosts  $C,\oC, \x $ one spinor $\f, \ov \f$, and a mass factor with some power of m, since they are not really anything but constants for present purposes.  Recall that we have set m=0 for simplicity. 
  We also note that the mass m and the spinors 
 $\f_{\dot \a},  \ov \f_{\a}$ do not transform at all.
  
\subsection{\Large  Terms with dimension = - 2 }

\refstepcounter{orange}
{ \theorange}.\; The   $\rm dimension =  \fr{1}{2}$   of $\f_{\dot \a}$  is chosen with a view to the \cdss\ which will be discussed in a future paper. It is a constant \cdss.
The lowest dimension  sectors that exist will be looked at in this section.
These are very small sectors, but they are easy to look at and they teach us something.  Here we will restrict ourselves to completely contracted expressions with low dimensions.  The only things that exist here are:
  \be
  \Ct  (\ov \f C) \A\in E_2, \oCt  (\f \oC)  A \in E_2
\ee
We cannot make any objects other than these here in $E_2$, given the discussion in section \ref{operatorchapter}.  Thus for example we do not need to consider $ \Ct  (\ov \f C) (C \y^a)$ because it contains a contraction 
$ (C \y^a)$, which is forbidden by the rules in section  \ref{operatorchapter}.
What this really means of course is that the mapping  $ \Ct  (\ov \f C) A^a$ $ \stackrel{d_1}{\lra}   \Ct  (\ov \f C) (C \y^a)$ is automatically taken into account when we restrict ourselves as suggested in section \ref{operatorchapter}.

The objects
$  \Ct$ and   $\oCt$ have dimension = - 3 and so the above have dimension = - 2.
These are already in $E_{\infty}$, and here is what they correspond to
  \be
  \Ct  (\ov \f C) \A_a \in E_{\infty} \ra \int d^4 x \oF_a \ov \f^{\a}  C_{\a} 
 \in \cH
  \ee

  \be
  \oCt  (\f \oC)  A^a  \in E_{\infty} \ra \int d^4 x F^a   \f^{\dot \a}  \oC_{\dot \a} 
 \in \cH
\ee

\refstepcounter{orange}
{ \theorange}.\;It is easy to see that the objects here  are in $\cH$ by looking at the table (\ref{brstransWZ}).  They are cocycles and not coboundaries.  From the spectral sequence point of view, there are only two objects here, and they have the same grading, so no possible $d_r, r\geq 2$ can kill either of them.

\subsection{\Large  Terms with dimension = - 1 }

 \refstepcounter{orange}
{ \theorange}.\;We will use the results of sections  \ref{fromdonefore2}
and \ref{explanationofdoneande2} here to jump directly to the space $E_2$. 
  Then we see that the only terms that exist here, in $E_2$,  are:
\be
 e_a \Ct  (\ov \f \y^a) ,     \Ct  (\ov \f C) \ov e^{bc}\A_b\A_c,  \eb
  \ov e^a  \oCt  ( \f \oy) ,   
  e_{bc}  \oCt  (\f \oC)  A^b A^c 
\ee
These do have mappings 
\be
e_a  \Ct  (\ov \f \y^a)  \stackrel{d_2}{\lra}  e_a \Ct  (\ov \f C) g^{abc}\A_b \A_c
  \ee 
  \be
  \ov e^a  \oCt  ( \f \oy_a)     \stackrel{d_2}{\lra}  
   \ov e^a g_{abc}  \oCt  (\f \oC)  A^b A^c 
\ee
and their adjoints
\be
\ov e^{bc}  g_{abc} \Ct  (\ov \f \y^a)  \stackrel{d_2^{\dag}}{\lla}   \Ct  (\ov \f C) \ov e^{bc}\A_b \A_c
  \ee 
  \be
  e_{bc}  \og^{abc}   \oCt  ( \f \oy_a)   \stackrel{d_2^{\dag}}{\lla} 
  e_{bc}  \oCt  (\f \oC)  A^b A^c 
\ee
This operator $d_2$ arises from:
\[
d_2= \P_2 \d_1 \d_0^{\dag} \d_1\P_2+*
 = \P_2 ( \og^{abc} \A_b\A_c ) \ov\Lam^{a \dag}) (F^d \ov\Lam^{d \dag} )^{\dag} 
 (F^e C_{\b} \y_{\b}^{e\dag})  
\P_2+*\]\be\equiv \P_2   ( \og^{abc} \A_b\A_c 
  C_{\a}) \y^{a \dag}_{\a} 
\P_2+*
\la{originofgoodd2}
\ee

\refstepcounter{orange}
{ \theorange}.\;There are no further $d_r$ 	because there is nothing left for it to operate on in this sector. The isomorphisms are

\be
e_a  \Ct  (\ov \f \y^a) \in E_{\infty}\ra    \int d^4 x  e_{a}   \ov \G^a \ov \f^{\a} C_{\a} \in \cH 
\ee
\be
  \ov e^a  \oCt  ( \f \oy_a)    \in E_{\infty}\ra  \int d^4 x  \ov e^{a}  \G_a   \f^{\dot \a}\oC_{\dot \a}\in \cH 
\ee

\be
  \Ct  (\ov \f C) \ov e^{bc}\A_b \A_c \in E_{\infty}\ra  \int d^4 x \ov e^{(ab)} 
\lt (  \A_a \oF_b - \fr{1}{2}\oy^{\dot \a}_a \oy_{b\dot \a} \rt )  (\ov \f C)
\in \cH 
\ee

\be
  e_{bc}  \oCt  (\f \oC)  A^b A^c 
 \in E_{\infty}\ra  \int d^4 x e_{(ab)} 
\lt ( \A^a F^b - \fr{1}{2}\y^{a \a} \y^{b}_{ \a} \rt )  (\f \oC) 
\in \cH 
\ee
 These are for tensors that satisfy the constraint equations, of course. Once again, a careful analysis of the constraint equations is needed here.

\section{\LARGE Terms with dimension = 0, zero spin,  and a constant spinor with $N_{\f} + N_{ \ov \f}=1 $}
\la{dimzerosec}

\subsection{\Large $N_{\x}=1$ Terms with dimension 0 }

\refstepcounter{orange}
{ \theorange}.\;Now we have a new kind of term:
 
\be
(C \x \f) \A_a,  (\oC \x \ov \f) \A_a, \eb
(C \x \f) A^a,(\oC \x \ov \f) A^a, \eb
(\oC \x \ov\f) (\oC\oy), 
(C \x \f) (C\y)\eb  
 (C \x \oC) (\f\oy),  
(C \x \oC) (\ov \f \y)  
\ee

Here are some relevant mappings:

\be
(C \x \f) A^a 
 \stackrel{d_1}{\lra}
(C \x \f) (C\y)
\ee
\be
(\oC \x \ov \f) \A_a 
\stackrel{d_1}{\lra}
(\oC \x \ov\f) (\oC\oy), 
\ee
\be
(C \x \f) \A_a \stackrel{d_1}{\lra}
 (C \x \oC) (\f\oy),  
\ee
\be
 (\oC \x \ov \f) A^a\stackrel{d_1}{\lra} 
(C \x \oC) (\ov \f \y)  
\ee

There are no survivors here in $E_2$.  Note that the cohomological structure of the   structure operator  is essential here. 

\subsection{\Large $N_{\x}=2$ Terms with dimension 0 }

\refstepcounter{orange}
{ \theorange}.\;Now we can write down the $  \Ct$ and   $\oCt$  terms in  $E_2$, given the discussion in section \ref{operatorchapter}.  The objects
$  \Ct$ and   $\oCt$ have dimension = - 3.

\be   \Ct  (\ov \f \y) \A,     \Ct  (\ov \f C) \A \A \A,    \oCt  ( \f \oy) A,\oCt  (\f \oC)  AAA
   \ee

These do have mappings, using the same operator that appeared earlier:
\be
e_{a}^{b}  \Ct  (\ov \f \y^a) \A_b \stackrel{d_2}{\lra}e_{a}^{b}  \Ct  (\ov \f C) \og^{adf}\A_d \A_f\A_b
  \ee 
  \be
  e_{a}^{b}    \oCt  ( \f \oy_b) A^a     \stackrel{d_2}{\lra}  
 e_{a}^{b}     g_{bde}  \oCt  (\f \oC)  A^d A^e A^a 
\la{constraintfirst}
\ee
and their adjoints
\be
e^{dfb} g_{afb}  \Ct  (\ov \f \y^a) \A_d \stackrel{d_2^{\dag}}{\lla}  \Ct  (\ov \f C) e^{dfb}\A_d \A_f\A_b
\la{constraintsecond}
  \ee 
  \be
 e_{ade} \og^{bde}   \oCt  ( \f \oy_b) A^a     \stackrel{d_2^{\dag}}{\lla}
      e_{ade}  \oCt  (\f \oC)  A^d A^e A^a 
\ee 
There are no further $d_r$.

\subsection{\Large Isomorphisms for dimension = 0 and $N_{\f}+ N_{\ov \f}=1$: the exotic pair $\cE,\W$ }
\la{wildone}

Here is $\cE$ for this case:
\be
 e^b_a\Ct  (\ov \f \y^a) \A_b \in E_{\infty} \ra   \ov \f^{\a}e^b_a \int d^4 x \lt \{  C_{\a}  \lt (\ov \G^a  \A_b+ \ov Y^{a\dot \b} \oy_{b \dot \b} + \ov \Lam^a \ov F_b  \rt )\ebp
+ \lt (  A^{a}\pa_{\a \dot \b} \oy_b^{\dot \b} 
+
 \y^{a}_{\a} \ov F_b \rt )
\rt \}\in \cH
\ee
Here is $\W$ for this case:
\be
  \Ct  (\ov \f C) e^{(dfb)}\A_d \A_f\A_b\in E_{\infty}  \eb
  \ra  \int d^4 x \ov e^{(abc)} 
\lt (  \A_a  \A_b \oF_c -  \A_a \oy^{\dot \a}_b \oy_{c\dot \a} \rt )  (\ov \f C)
\in \cH 
\ee
 The complex conjugates can be derived in the usual way.

 These are for tensors that satisfy the constraint equations, of course. Once again, a careful analysis of the constraint equations is needed here.

 \section{\LARGE Terms with dimension = 1, zero spin,  and a constant spinor with $N_{\f} + N_{ \ov \f}=1 $}
 \la{nightmare}

\refstepcounter{orange}
{ \theorange}.\; 
Here we will start by  treating  the terms with different values of $N_{\x}$, separately. Since $[N_{\x} , d_1]=0 $ this makes sense to determine what survives to $E_2$.  But then it can, and does,  happen that $d_r, r\geq 2 $ maps items in $E_2$ with different values of $N_{\x}$.

\subsection{\Large $\rm N_{\x}=0$ ,Terms with dimension = 1, zero spin,  and a constant spinor with $N_{\f} + N_{ \ov \f}=1 $ with just one field}
\refstepcounter{orange}
{ \theorange}.\; There are six types of terms here:
    \be
 (C \ov \f) A^a ,(\oC  \f) \A_a , \eb
(\oC  \f) A^a,   (C \ov \f) \A_a  ,\eb
 (C \ov \f) (C \y^a ),(\oC  \f) (\oC \ov \y_a)
\ee  
\refstepcounter{orange}
{ \theorange}.\;    We note the following maps by the operator $d_1$ in the  $N_{\x}=0$ sector:
\be
(C \ov \f) A^a  \stackrel{d_1}{\lra} (C \ov \f) (C \y^a )
\ee
\be
  (\oC  \f) \A_a   \stackrel{d_1}{\lra} (\oC  \f) (\oC \ov \y_a)
\ee
So these four terms do not survive to  $E_2$. 
However, because there are no $d_1$  mappings involving the following terms, the following two kinds of terms   do survive to $E_2$ in this  $N_{\x}=0$ sector:
 \be
 (\oC  \f) A^a \in E_2; 
   (C \ov \f) \A_a  \in E_2
\la{tobekilled}   \ee  
   These do not get killed by $d_1$ because there are no terms
 $C \oC$ in the cohomology space $E_1$ in the $N_{\x}=0$ sector. But we will see below that these do get killed in the $E_2$ space by a form of $d_2$.

\subsection{\Large $\rm N_{\x}=1$ ,Terms with dimension = 1, zero spin,  and a constant spinor with $N_{\f} + N_{ \ov \f}=1 $  with just one field}

\refstepcounter{orange}
{ \theorange}.\; There are six types of terms here:
  \be
 (C \x)_{\dot \a}A^{a  \b \dot \a} {\ov \f}_{\b} \in E_1
,   
 (\oC \x)_{  \b}\A_{a}^{ \dot \a \b}
 { \f}_{\dot \a} \in E_1 
 \ee
   \be
 (C \x)_{\dot \a}\y^{a (\a \b)  \dot \a} C_{\a} {\ov \f}_{\b} \in E_1
,   
 (\oC \x)^{  \d}  \oy_{a(\dot \a\dot \b)  \d} 
 { \f}^{\dot \a} \oC^{\dot \b} \in E_1 
 \ee
  \be
 (C \x)_{\dot \a}\A_a^{\dot \a \b } {\ov \f}_{\b} \in E_1
,   
 (\oC \x)_{  \b}A^{a \b \dot \a}
 { \f}_{\dot \a} \in E_1 
 \ee
 
\refstepcounter{orange}
{ \theorange}.\; 
The first four terms here get mapped simply by $d_1$ as follows:
  \be
 (C \x)_{\dot \a}A^{a  \b \dot \a} {\ov \f}_{\b} \stackrel{d_1}{\lra}
 (C \x)_{\dot \a}\y^{a (\a \b)  \dot \a} C_{\a} {\ov \f}_{\b}  
\ee
\be
 (\oC \x)_{  \b}\A_{a}^{ \dot \a \b}
 { \f}_{\dot \a} \stackrel{d_1}{\lra}
 (\oC \x)^{  \d}  \oy_{a(\dot \a\dot \b)  \d} 
 { \f}^{\dot \a} \oC^{\dot \b} \ee

\refstepcounter{orange}
{ \theorange}.\; 
The last two terms do not get killed in $E_1$ because they map to a term which vanishes by the equation of motion, as we will see. 
Thus we have
 \be
 (C \x)_{\dot \a}\A_a^{\dot \a \b } {\ov \f}_{\b}   \stackrel{d_1}{\lra}
(C \x \oC) \oy_a^{(\dot \a  \dot \b) \b } \ve_{\dot \a  \dot \b}
 {\ov \f}_{\b}= 0
\ee
where the latter equation arises because 
\be
(C \x \oC) \oy_a^{(\dot \a  \dot \b) \b } \ve_{\dot \a  \dot \b}
 \not \in E_1 
 \ee
 since $\oy_a^{(\dot \a  \dot \b) \b } \ve_{\dot \a  \dot \b}$ is not a symmetric variable. So it follows that
 \be
 (C \x)_{\dot \a}\A_a^{\dot \a \b } {\ov \f}_{\b} \in E_2
,   
 (\oC \x)_{  \b}A^{a \b \dot \a}
 { \f}_{\dot \a} \in E_2
 \la{tobekilled2}
 \ee
These get mapped together with the terms in (\ref{tobekilled}) as follows\footnote{Here we use $d_2 = \P_2 \d_2 \P_2 $ with (\ref{extderd2}) for the differential.}

\be
(C  \ov \f) \A_a   \stackrel{d_2}{\lra}  (C \x)_{\dot \a}\A_a^{\dot \a \b }
 {\ov \f}_{\b}   
\ee
and
\be
(\oC  \f) A^a \stackrel{d_2}{\lra}  (\oC \x)_{ \a} A^{a \a \dot \b}
 { \f}_{\dot \b}   
 \ee
 and so none of these four terms  (\ref{tobekilled}) and  (\ref{tobekilled2}) survive to $E_3$. 
 
  So none of the terms with just one field here survive to $E_3$.
 Now we turn to terms with two fields.

\subsection{\Large $\rm N_{\x}=1$ Terms with dimension = 1, zero spin,  and a constant spinor with $N_{\f} + N_{ \ov \f}=1 $:  Two Field   $(C\x) C\cdots C$ type Terms}
  \la{twofielddim1xiCterms}

Here is  a first set of  Two Field   $(C\x) C\cdots C$ type Terms in $E_1$:
 \be
(C \x \oy_a) (\ov \f C) A^b ,(C \x)^{\dot \d}  \A_{a \g \dot \d} C^{\d} (\ov \f C)A^b,
\la{yipe1}\eb
(C \x \oy_a) (\ov \f C) (C \y^b ), (C \x)^{\dot \d}  \A_{a \g \dot \d} C^{\d} (\ov \f C) (C \y^b )
\la{yipe2} \ee
 The following kills all the terms in lines
(\ref{yipe1}) and (\ref{yipe2}):
 \be
e_b^a
(C \x \oy_a) (\ov \f C) A^b  \stackrel{d_1}{\lra} 
e_b^a(C \x)^{\dot \d}  \A_{a \g \dot \d} C^{\d}(\ov \f C)A^b \eb
\oplus e_b^a (C \x \oy_a) (\ov \f C) (C \y^b ) 
  \stackrel{d_1}{\lra}
  e_b^a (C \x)^{\dot \d}  \A_{a \g \dot \d} C^{\d} (\ov \f C) (C \y^b )
\ee
Next consider the second set:
 \be
e^{ab}
(C \x \oy_a) (\ov \f C) \A_b 
;\;
 e^{ab}(C \x)^{\dot \d}  \A_{a \g \dot \d} C^{\d} (\ov \f C) \A_b 
\la{yipe3}\ee
The following kills the terms in line
(\ref{yipe3}):
 \be
e^{ab} (C \x \oy_a) (\ov \f C) \A_b  \stackrel{d_1}{\lra} e^{ab} (C \x)^{\dot \d}  \A_{a \g \dot \d} C^{\d} (\ov \f C) \A_b 
 \ee
Next consider the third set:
\be
e_{ab}(C \x)^{\dot \d}  A^a_{ \g \dot \d} C^{\g} (\ov \f C) A^b , 
e_{ab}(C \x)^{\dot \d}  \y^a_{(\e \g), \dot \g}C^{\e} C^{\g} (\ov \f C) A^b ,\eb
e_{ab}(C \x)^{\dot \d}  A^a_{ \g \dot \d} C^{\g} (\ov \f C) (\y^b C), 
e_{ab}(C \x)^{\dot \d}  \y^a_{(\e \g), \dot \g}C^{\e} C^{\g} (\ov \f C) (\y^b C) 
\ee

Here is the mapping for the third set:
\be
e_{ab}(C \x)^{\dot \d}  A^a_{ \g \dot \d} C^{\g} (\ov \f C) A^b 
 \stackrel{d_1}{\lra} \eb
e_{ab}(C \x)^{\dot \d}  \y^a_{(\e \g), \dot \g}C^{\e} C^{\g} (\ov \f C) A^b
 \oplus
e_{ab}(C \x)^{\dot \d}  A^a_{ \g \dot \d} C^{\g} (\ov \f C) (\y^b C) \stackrel{d_1}{\lra} \eb
e_{ab}(C \x)^{\dot \d}  \y^a_{(\e \g), \dot \g}C^{\e} C^{\g} (\ov \f C) (\y^b C) \ee
The notation $\oplus$ indicates that two different linear combinations are needed for the two mappings.  None of these survive to $E_2$.
Here is the fourth set
\be
  e_{a}^{b}
  (C \x)^{\dot \d}  A^a_{ \g \dot \d} C^{\g} (\ov \f C) \A_b   
  ,\eb
e_{a}^{b}
(C \x)^{\dot \d}  \y^a_{(\e \g), \dot \g}C^{\e} C^{\g} (\ov \f C)  \A_b  
\la{ohmy3}
\ee
and here is the mapping for the fourth set
\be
  e_{a}^{b}
  (C \x)^{\dot \d}  A^a_{ \g \dot \d} C^{\g} (\ov \f C) \A_b   
\stackrel{d_1}{\lra} 
e_{a}^{b}
(C \x)^{\dot \d}  \y^a_{(\e \g), \dot \g}C^{\e} C^{\g} (\ov \f C)  \A_b  
\la{ohmy2}
\ee
Neither of these survive to $E_2$.
Here is the fifth set:
\be
(C \x \f) \A_a \A_b,(C \x \oC) (\f\oy_a)\A_b 
\la{ego1}
\ee
The following kills the symmetric terms in line
(\ref{ego1}):
\be
(C \x \f) e_{(ab)}\A_a \A_b \stackrel{d_1}{\lra}
 (C \x \oC) e_{(ab)}(\f\oy_a)  \A_b 
\la{symfromeeqo1}
\ee
But the antisymmetric term does not get killed:
\be
 (C \x \oC) e_{[ab]}(\f\oy_a)  \A_b \in E_2
\la{antisymnotdead}
\ee
Then there is a sixth set:
\be
e_b^a (C \x \f) \A_a A^b,    e_b^a (C \x \oC) ( \f \oy^b) A^a, e_b^a (C \x \f) (C\y^b)\A_a
\la{ego2}
\ee
The following kills two out of the  three terms  in line
(\ref{ego2}):
\be
e_b^a(C \x \f) A^b \A_a 
 \stackrel{d_1}{\lra}
e_b^a(C \x \f) (C\y^b)\A_a +e_b^a  (C \x \oC) (\f\oy_a)A^b
\ee
But there is a linear combination that does not get killed. We denote this by:
\be
e_b^a(C \x \f) (C\y^b)\A_a \oplus e_b^a  (C \x \oC) (\f\oy_a)A\in E_2
\la{linex1f}\ee

Here is the seventh set:
\be
(C \x \f) A^a A^b, (C \x \f) A^a(C\y^b) , (C \x \f) (C\y^a)(C\y^b)  
\la{ego3}
\ee
The following kills all three terms, with both symmetries, in line
(\ref{ego3}):
\be
(C \x \f)
e_{(ab)}
 A^a A^b
\stackrel{d_1}{\lra} 
e_{(ab)}(C \x \f) A^a(C\y^b) 
\ee
\be
e_{[ab]}
(C \x \f) A^a(C\y^b) 
\stackrel{d_1}{\lra} 
e_{[ab]}
 (C \x \f) (C\y^a)(C\y^b)  
\ee
 None of these survive to $E_2$.

\subsection{\Large $\rm N_{\x}=1$ Terms with dimension = 1, zero spin,  and a constant spinor with $N_{\f} + N_{ \ov \f}=1 $:  Two Field   $(\oC\x) \oC\cdots \oC$ type Terms}
These are the \CC s of the above terms in section \ref{twofielddim1xiCterms}.  We will not repeat them all.  But we will take some of them here:
  Consider the terms:
 \be
  (\oC \x \ov \f) \A_a \A_b, (\oC \x \ov\f) (\oC\oy)\A_a, (\oC \x \ov\f) (\oC\oy)(\oC\oy)
\la{cego1}
\eb
   (\oC \x \ov \f) \A_a  A^b,  (C \x \oC) (\ov \f \y^b)\A_a ,
    (\oC \x \ov\f) (\oC\oy) A^a
\la{cego2}
\eb
 (\oC \x \ov \f) A^a A^b ,(C\x \oC) (\y^a\ov \f)  A^b 
\la{cego3}\ee
The following kills both terms, with both symmetries, in line
(\ref{cego1}):
\be
\ov e^{(ab)}
(\oC \x \ov \f) \A_a \A_b
\stackrel{d_1}{\lra}
\ov e^{(ab)} (\oC \x \ov\f) (\oC\oy_a)\A_b  
\ee
\be
\ov e^{[ab]}
(\oC \x \ov\f) (\oC\oy)\A_b \stackrel{d_1}{\lra}\ov e^{[ab]}
 (C \x \f) (C\y^a)(C\y^b)  
\ee
The following kills two out of three terms in line
(\ref{cego2}):
\be
(\oC \x \ov \f) \A_a A^b
\stackrel{d_1}{\lra}
(\oC \x \ov\f) (\oC\oy_a)A^b +(C \x \oC) (\ov \f \y^b)  \A_a, 
\ee
but there is a linear term which is not killed
\be
(\oC \x \ov\f) (\oC\oy_a)A^b \oplus (C \x \oC) (\ov \f \y^b)  \A_a, 
\la{linex2f}
\ee
The following kills the symmetric  terms in line
(\ref{cego3}):
\be
e_{(ab)} (\oC \x \ov \f) A^a A^b\stackrel{d_1}{\lra} 
(C \x \oC)e_{(ab)}   (\ov \f \y^a)  A^b
\la{symfromceqo3}
\ee
But the antisymmetric term does not get killed:
\be
(C \x \oC)e_{[ab]}   (\ov \f \y^a)  A^b\in E_2
\la{antisymnotdeadcc}
\ee

The rest of the complex conjugates are easy to find from the above results.

\subsection{\Large $\rm N_{\x}=1$ Terms with dimension = 1, zero spin,  and a constant spinor with $N_{\f} + N_{ \ov \f}=1 $:  Two Field    $(C\x\oC)$ type Terms}
These have the form

\be
e_b^a  (C \x \oC) (\f\oy_a)A^b \in E_1
;e_a^b  (C \x \oC) (\ov \f\y^a)\A_b\in E_1
\ee
These appear on the right in \ref{linex1f}:
\be
e_b^a(C \x \f) (C\y^b)\A_a \oplus e_b^a  (C \x \oC) (\f\oy_a)A^b\in E_2
\la{getskilled}
\ee
and on the right in \ref{linex2f}:
\be
e_b^a(\oC \x \ov\f) (\oC\oy_a)A^b \oplus e_b^a (C \x \oC) (\ov \f \y^b)  \A_a\in E_2
\la{getskilledcc}
\ee
Then there are also two terms of the following form, which survive from 
(\ref{antisymnotdead})
and
(\ref{antisymnotdeadcc}):
\be
 (C \x \oC) \ov e^{[ab]}(\f\oy_a)  \A_b  \in E_2
;(C \x \oC) e_{[ab]}   (\ov \f \y^a)  A^b  \in E_2
\la{antisymmstillherine2}
\ee

\subsection{\Large $\rm N_{\x}=2$ Terms in $E_2$ with dimension = 1, zero spin,  and a constant spinor with $N_{\f} + N_{ \ov \f}=1 $:  Two Field    $\Ct$ and $\oCt$  type Terms}
Here we can and do use the criteria in section \ref{operatorchapter} to restrict this to just a few terms in $E_2$.  They are as follows:

 These are the terms bilinear in fields
 \be
e_b^a  \Ct  \f^{\dot \b} \A_{a\dot \b \a } \y^{b\a}\in E_2
\la{bilincTterm}\ee
\be
\ov e^b_a \oCt  \ov \f^{ \a} A^a_{\a \dot \b} \oy_b^{\dot \b}\in E_2
\la{bilinocTterm}\ee

 These are the terms trilinear in fields
 \be e_{ab}^c \oCt  ( \f \oy_c) A^a  A^b \in E_2; \ov  e^{ab}_c   \Ct  (\ov \f \y^c)\A_a \A_b \in E_2 ;
  \ee

 These are the terms quadrilinear in fields
 
\be
   e_{abde}   \oCt  (\f \oC) A^a  A^b A^d  A^e \in E_2;\ov e^{abde} 
    \Ct  (\ov \f C) \A_a \A_b \A_d \A_e\in E_2
\ee

Those are all we can make, given the results of section \ref{operatorchapter}. 

\subsection{\Large Mappings involving the $\rm N_{\x}=2$ Terms with dimension = 1, zero spin,  and a constant spinor with $N_{\f} + N_{ \ov \f}=1 $:  Two Field    $\Ct$ and $\oCt$  type Terms}

\refstepcounter{orange}
{ \theorange}.\; 
Now we note the mapping:
\be
e_b^a(C \x \f) (C\y^b)\A_a  \in E_2 \stackrel{d_2}{\lra}  
e_b^a  \Ct  \f^{\dot \b} \A_{a\dot \b \a } \y^{b\a}\in E_2
\ee
\be
\ov e_b^a (\oC \x \ov\f) (\oC\oy_a)A^b  \in E_2\stackrel{d_2}{\lra}   \ov e^b_a \oCt  \ov \f^{ \a} A^a_{\a \dot \b} \oy_b^{\dot \b}\in E_2
\ee
which removes all four of those terms exactly and so this removes the terms 
(\ref{getskilled}) and
(\ref{getskilledcc}), as well as  the two terms (\ref{bilincTterm}) and  {\ref{bilinocTterm}})

\refstepcounter{orange}
{ \theorange}.\; 
  However we now get
\be e_{ab}^c \oCt  ( \f \oy_c) A^a  A^b \stackrel{d_2}{\lra} 
    e_{ab}^c  g_{cde} \oCt  (\f \oC) A^a  A^b A^d  A^e\ee
\be
  \ov  e^{ab}_c   \Ct  (\ov \f \y^c)\A_a \A_b  \stackrel{d_2}{\lra}
   \ov  e^{ab}_c  \og^{cde}\Ct  (\ov \f C) \A_a \A_b \A_d \A_d
\ee
which is a constraint like the ones we have seen before.

\refstepcounter{orange}
{ \theorange}.\; 
But we must not forget that are still two terms of the following form, which survive from 
(\ref{antisymmstillherine2}).  In fact, they do not get mapped by $d_2$ either, so they live to $E_3$: 
\be
 (C \x \oC) \ov e^{[ab]}(\f\oy_a)  \A_b  \in E_3
;(C \x \oC) e_{[ab]}   (\ov \f \y^a)  A^b  \in E_3
\ee 

 and then we get another constraint on the solutions of the above constraints:
\be
 (C \x \oC) \ov e^{[ab]}(\f\oy_a)  \A_b  \in E_3\stackrel{d_3}{\lra}
\ov  e^{[ab]} g_{bcd} \oCt  ( \f \oy_a) A^c  A^d  \in E_3
\ee
\be
(C \x \oC) e_{[ab]}   (\ov \f \y^a)  A^b  \in E_3\stackrel{d_3}{\lra}
 e_{[ab]} \og^{bcd}     \Ct  (\ov \f \y^a) \A_c \A_d\in E_3
\ee
See the discussion above in paragraph \ref{d3paragraph} with respect to this version of $d_3$.  We also have the adjoint constraints of course:
\be 
    e_{abde}  \oCt  (\f \oC) A^a  A^b A^d  A^e
     \stackrel{d_2^{\dag}}{\lra} 
      e_{abde} \og^{cde} \oCt  ( \f \oy_c) A^a  A^b
     \ee
\be
  \ov  e^{abde}    \Ct  (\ov \f C) \A_a \A_b \A_d \A_d
      \stackrel{d_2^{\dag}}{\lra}
 \ov  e^{abde}  g_{cde} \Ct  (\ov \f \y^c)\A_a \A_b 
\ee

and
\be
   e^{a}_{cd}  \oCt  ( \f \oy_a) A^c  A^d   
 \stackrel{d_3^{\dag}}{\lra}
 (C \x \oC)   e^{[a}_{cd} \og^{b] cd}  (\f\oy_a)  \A_b   
\ee

\be
\ov e_{a}^{cd}     \Ct  (\ov \f \y^a) \A_c \A_d
\stackrel{d_3^{\dag}}{\lra}
(C \x \oC) \ov e_{[a}^{cd}  g_{b] cd} (\ov \f \y^a)  A^b  
\ee

\refstepcounter{orange}
{ \theorange}.\; Now we note that
\be
E_4 = E_{\infty}\ra \cH
\ee
because there is nothing left for a higher $d_r, r\geq 4$ to operate on. We only have terms left here with $\D N_{\rm Grading}=0$ and $\D N_{\rm Grading}=3$.

\subsection{\Large Isomorphisms for dimension = 1 and $N_{\f}+ N_{\ov \f}=1$: the exotic triplet $\cC,\cE,\W$}
\la{thecCinvariant}
\refstepcounter{orange}
{ \theorange}.\; 
We will indicate only the new kind of item here. Here is $\cC$ for this case.  Its existence is a confirmation that our rather long and complicated derivation above is correct:
\be
 (C \x \oC) \ov e^{[ab]}(\f\oy_a)  \A_b  \in E_{\infty} \ra \eb
\ov e^{[ab]} \int d^4 x \lt \{
 \G_a (\f \ov \y_b)+ Y_{a\a} \f_{\dot \b} \pa^{\a \dot \b} \A_b+ 
\lt ( \Lam_a     \G_b - \fr{1}{2}
  Y_{a}^{\a} Y_{b \a} \rt ) (\f \oC)
 \rt \}\in \cH
\ee
Note that this term has ghost charge minus one. 
Observe the strange double pseudofield terms $\ov e^{[ab]}\lt ( \Lam_a     \G_b - \fr{1}{2}
  Y_{a}^{\a} Y_{b \a} \rt ) (\f \oC)$. 
This isomorphism assumes that the constraint equations are satisfied, of course.  The other two isomorphisms for $\cE$ and $\W$ are similar to those in section \ref{wildone} above.   These, as well as higher examples of dimension and spin, will be left for further research.  Judging from this section, that will involve considerable work.

\section{\LARGE Summary of the Construction of $E_{\infty}\ra \cH$:}
\la{sumsec}
\refstepcounter{orange}
{ \theorange}.\; Now that we have seen plenty of examples,  we will try to summarize the process of constructing this spectral sequence.  This should help with understanding how it works, and also should help with an understanding of how to apply it to  spins and dimensions  beyond those in this paper. Here we assume that the grading is as shown in (\ref{Ngrading}) and that the operator $\d$ is as shown in (\ref{brstransWZ}), and that the methods of \ci{dixspecseq} and the results of \ci{dixmin} are used. To start with, one must choose a dimension and  decide whether to  include the constants $\f,\ov \f$ or not, and then look for terms of zero spin. 

\subsection{\LARGE Description of $E_1$:}
\la{summaryofcontructionofE1}
\refstepcounter{orange}
{ \theorange}.\; First of all it is necessary to grasp what $E_1$ looks like.  It is composed of expressions made from the symmetrized fields and the cohomology of $\d_{\rm Structure}$, all as discussed in section \ref{formofE1para}.
For the $\Ct$ sector, we can choose to use the results of section \ref{operatorchapter}.
These are as follows:
\ben
\item
We use the symmetrized fields to construct the coefficient $T$ for $\Ct$.
This depends on the dimension chosen, and of course the dimension of $\Ct$ 
is dimension = - 3.
\item
If there is no C present, so that the ghost charge of $\Ct T$   is zero (Form charge 4), we can use any of the symmetrized fields to construct $T \in E_1$, but $T$ must have spin zero of course. 
\item
In addition, assuming there is no extra C present, the expression must satisfy the constraint
\be
\na_{\a} T=0
\la{constraintforT}
\ee
Since $\na_{\a}$ has the form \ref{formofna}, this can be satisfied easily by ensuring that $T$ does not contain any fields $\oy$ or $A$.  But even if it does contain some fields $\oy$ or $A$, there are plenty of solutions for (\ref{constraintforT}).
\item 
If there are one or more extra  C's present, so that the ghost charge of $\Ct T^{\a_1 \cdots \a_1}
 C_{\a_1} \cdots C_{\a_n} $  is n   (Form charge 4+n),  we  have more restrictions:
 \ben
 \item
We can use only the symmetrized fields $\y$ and $\A$ to construct $T \in E_1$
\item
$T$ must have spin zero, of course. 
\item
Any undotted indices on the symmetrized fields  $\y$ and $\A$ must appear in an uncontracted symmetrized form with the indices on  the ghosts $C_{\a_1} \cdots C_{\a_n}$. 
\item
No extra requirement of the form (\ref{constraintforT}) is needed since that equation is automatically satisfied with these variables. 
\item
Any number of the underived fields $\A_a$ can be used since there is no need to symmetrize its undotted indices  with $C_{\a_1} \cdots C_{\a_n}$, since it has no undotted indices.
\een
\een
The $\oCt$ sector is the \CC\ of the $\Ct$ sector.

\subsection{\Large Application  of $d_1$ and Description of $E_2$:} 
\la{applicationofd1andlistofE2}
\refstepcounter{orange}
{ \theorange}.\; Here we assume that we have a complete set of terms as constructed above in section \ref{summaryofcontructionofE1}. At this point we need to apply the operator $d_1$ to the various terms and see which ones are mapped to each other. Some familiarity with the operators is useful here.   Both ends of the mapping lead to an exclusion from $E_2$. After all of these mappings have been found, we must list all the terms that have survived to $E_2$.

\subsection{\Large Application  of $d_2$ and Description of $E_3$:} 
\refstepcounter{orange}
{ \theorange}.\; This requires us to observe the list of the terms that have survived to $E_2$ after observing the maps in section \ref{applicationofd1andlistofE2} above.
These can become involved in maps from $d_2$.  At this point constraint equations of various kinds naturally emerge in some cases.  There are a number of ways that this differential can arise:
\ben
\item
The simplest example is 
\be
d_2 = \P_2 \x \pa \P_2
\ee
No constraint equations emerge from this, because it does not involve the superpotential coefficients. 
\item
There are a number of possibilities next. One kind has the form of (\ref{weirdd2}). 
It gives no constraint equations. 
\item
Another is in (\ref{originofgoodd2}).
\item
This seems to be fairly simple, because by this point there is not much left in the space.  One simply looks for the relevant values of $N_{\rm Grading}$, and then searches for the relevant $d_r$ using the formulae and explanations in \ci{dixspecseq}.
\een

 \subsection{\Large   Higher Differentials $d_r,r\geq 3$ and Description of $E_{r+1},r\geq 3$:} 
\refstepcounter{orange}
{ \theorange}.\; This also seems to be fairly simple, because by this point there is not much left in the space.  One simply looks for the relevant values of $N_{\rm Grading}$.
The theory typically generates constraint equations.  Solving them is the next problem.

\section{\LARGE Conclusion}
\la{conclussec}
\refstepcounter{orange}
{ \theorange}.\;In this paper we have shown how to calculate the full BRS cohomology for the Wess Zumino model, including the pseudofield sources, using the spectral sequence method.  It must be admitted that this paper is long and complicated.  However that seems rather inevitable, given the experience that mathematicians have had with the spectral sequence in other circumstances, as evidenced by the quotes in section \ref{quotesaboutspecseq}.  Also, when we look at the results in sections \ref{dimzerosec} and \ref{nightmare}, it is clear that the way the \ED\ works is really quite obscure and surprising, and so are the results for $\cH$.

\refstepcounter{orange}
{ \theorange}.\; 
The improvement here is that the previous effort in \ci{dixminram} did not include the pseudofield sources, and so it missed the invariants that can be anomalous.  These are all made with the pseudofields.  The simplest of these, for the interacting but massless theory is in  section \ref{wildone}.  It is the following member of the exotic pair $(\cE,\W)$: 
\be
 \Ct e^b_a   (\ov \f \y^a) \A_b \in E_{\infty} \ra \cE=
 e^b_a  \ov \f^{\a}  \int d^4 x  \eb
 \lt \{  C_{\a}  \lt (\ov \G^a  \A_b+ \ov Y^{a\dot \b} \oy_{b \dot \b} + \ov \Lam^a \ov F_b  \rt )
+   \lt (  A^{a}\pa_{\a \dot \b} \oy_b^{\dot \b} 
+
 \y^{a}_{\a} \ov F_b \rt )
\rt \}\in \cH
\la{funnynewinvariant}
\ee 
Clearly this depends crucially on the presence of the pseudofields $\ov \G^a  , \ov Y^{a\dot \b}, \ov \Lam^a$ and could not exist without them. Several more strange invariants arise in section \ref{thecCinvariant}, including the following member of the exotic triplet  $(\cC,\cE,\W)$: 
\be
 (C \x \oC) \ov e^{[ab]}(\f\oy_a)  \A_b  \in E_{\infty} \ra \cC=\eb 
\ov e^{[ab]} \int d^4 x \lt \{
 \G_a (\f \ov \y_b)+ Y_{a\a} \f_{\dot \b} \pa^{\a \dot \b} \A_b+ 
\lt ( \Lam_a     \G_b - \fr{1}{2}
  Y_{a}^{\a} Y_{b \a} \rt ) (\f \oC)
 \rt \}\in \cH
\ee
  The discovery of these peculiar objects is a good argument that the spectral sequence method here is working well. It is quite easy to check that they are in the space $\cH$ just by using table (\ref{brstransWZ}), assuming the constraint equations are satisfied. But guessing them without the spectral sequence seems very challenging, and superspace would not help, because they are not supersymmetric. They are not scalars in  superspace.  They can only be written in  superspace with explicit factors of $\q$ and $\oq$ \ci{dixexotic}. 

The next steps in this research probably need to include:
\ben
\item
Solutions of the constraint equations like we saw in (\ref{constraintfirst}): 
\be
 g_{ d(bc}  e^{d}_{a)}=0
 \ee
 and their adjoints like we saw in (\ref{constraintsecond}) :
 \be
 g_{ dbc}   \ov e^{ abc} =0 
 \ee
\item
Calculation of the possible supersymmetry anomalies relating to the new BRS invariants and exotic pairs.  If we want to generate triangle diagrams to calculate the anomaly coefficients, it will be necessary to either use the dimension one case or else go to the case where the constant spinor becomes a full \cdss. In addition one needs to have a solution for the constraint equations in mind, and that is another topic too.   One could try to generate the anomaly with the dimension zero case, but that involves diagrams with only one momentum, and that is also fraught with complications.

\item
Introduction of the \cdss\ and its spectral sequence and its action and solution of the tachyon issue.
\item
Introduction of gauge fields and interactions, and spontaneous breaking of gauge symmetry.
\een

\refstepcounter{orange}
{ \theorange}.\;We have restricted the treatment  to low dimensions here.   To treat higher dimensions require us to analyze the   terms carefully, probably using the \ED, for each dimension $ = 1, 2\cdots$.  But, for now, the above results already present plenty of questions for further research.

{
\refstepcounter{orange}
{ \theorange}.\; The \cdss \ has an interesting action. 
In the paper \ci{dixexotic}, we did not attempt to discuss the basic action of the \cdss s, and we did not attempt to address the situation of massive theories.   One reason for that was that it was a concern that the simplest guess for the  action of a \cdss\ contains tachyonic degrees of freedom. In a paper being prepared it will be shown that an extended action, including a new kind of kinetic term, allows us to choose parameters to eliminate those tachyonic degrees of freedom.  This action will be incorporated into the spectral sequence, so that we can discuss the exotic pairs in a way that does not suffer from tachyonic problems, and we can also consider mass.

\refstepcounter{orange}
{ \theorange}.\;There is another new puzzling issue here too.  Normally when one  adds a new invariant to the action, because of renormalization for example, one  can add it as discussed in \ci{dixonnucphys}, which shows that  the new terms of higher orders can arise by renormalizing the invariants and also  introducing a canonical transformation.  This canonical transformation is generated by a generating functional $\cF$ that has ghost charge minus one, in accord with the  conjecture that 
\be
\d_{\rm BRS} \cA_{\rm Counterterms}=0 \Ra 
\cA_{\rm Counterterms}=\d_{\rm BRS} \cF + \cA_{\rm invariants}
\la{oldtheorem}
\ee
where the term $ \cA_{\rm invariants}$ depends only on the fields, and not on the pseudofields or the ghosts. There have been various attempts to prove  this conjecture in various theories.  Those attempts are not easy to understand, and they do not use the spectral sequence.  The results of the present paper indicate that the real situation is sometimes not consistent with this conjecture. Moreover, it seems likely that the complexity of the present results could not be obtained except by using the spectral sequence.

\refstepcounter{orange}
{ \theorange}.\;The conjecture (\ref{oldtheorem}) works well so long as the invariants depend only on the fields in the theory. 
But with these new kinds of invariants in the cohomology space like (\ref{funnynewinvariant}), this no longer makes sense, because instead of 
$\cF$ with ghost charge minus one, we are generating a term $\cA_{\rm invariants}$ that has ghost charge zero: yet it contains the \Pf s and the 
ghosts. 

\refstepcounter{orange}
{ \theorange}.\; The point, of course, is that the conjecture (\ref{oldtheorem}) is simply not true when we have terms like (\ref{funnynewinvariant}) in the cohomology space.   If we simply add the new invariants to the action that generates at least  two problems:
\ben
\item
The pseudofield terms generate new terms in the field transformations, so that a new cohomology problem emerges.
\item
It is not clear that the addition of the new invariants results in an action that is invariant under either the old or the new transformations. 
 \een

 \refstepcounter{orange}
{ \theorange}.\; So an attempt to generate a canonical transformation from the cohomology as in equation (3.4) of \ci{dixonnucphys} does not work, since $\d_{\rm BRS}$ changes when one adds the new invariant.  There are related issues that arise from the fact that the exotic invariants necessarily violate superspace invariance too, as mentioned in \ci{dixexotic}, so that the nonrenormalization theorems of chiral supersymmetry do not apply here. 
     These issues require analysis and thought, and the solution, and meaning,  of these puzzles is currently unknown.

 \section{\LARGE Glossary}
 \la{glossary}
 
{\bf Adjoint:}  We use the adjoint to denote the operator which satisfies   equations like the following for Grassmann Even bosons: 
\be
\lt [ A^{a \dag}, A^b \rt ]= \d^a_b
;\lt [ \A_{a}^{ \dag}, \A_b \rt ]= \d^b_a
\ee
 and the following for Grassmann Odd fermions:
\be
\lt \{ \y_{\a}^{a \dag}, \y_{\b}^b \rt \}= \d^b_a \d^{\a}_{\b}
; \lt \{ \oy_{\dot \a}^{a \dag}, \oy_{\dot \b}^b \rt \}= \d^b_a \d^{\dot\a}_{\dot \b}
\ee
 This is equivalent to defining the adjoint as the derivative with respect to the field as follows
 \be
A^{a \dag}\equiv \fr{\pa}{\pa A^a }
; \y_{\a}^{a \dag}\equiv \fr{\pa}{\pa \y_{\a}^{a } }; \; {\rm etc.}
\ee
These definitions generate a Fock space out of the local field polynomials, with a postive definite metric, as explained in \ci{dixspecseq}. This contains all possible derivatives too, for example
 \be
A^{a \dag}_{,\a \dot \b} \equiv \fr{\pa}{\pa A^{a}_{,\a \dot \b}}
\ee
Taking the adjoint of a product of fields and ghosts means also reversing the order of all operators and taking the complex conjugate of all numbers and the adjoint of all fields, but this does not mean taking the complex conjugate of complex fields.  We also do not include a minus sign when reversing the order of fermions while taking the adjoint of a product of two fermions. That leads to an unnecessary lack of symmetry under the operations of complex conjugation and taking the adjoint. Note that this adjoint is for the purpose of calculating cohomology, not for the purpose of showing that the action is hermitian.  See 
Complex Conjugation in this Glossary with regard to that issue.

{\bf Anomaly:}  In a quantum field theory, one expects that the infinities of the Feynman diagrams, computed using any sensible regularization method, will obey the symmetries of the theory.  That means, that if one makes a variation, according to the symmetry, in a Feynman calculation, that one  should obtain a finite, calculable, and local result, even though the original calculation gives, of course, a nonlocal and usually infinite result.  One can expect the infinities to cancel, if one  takes the variation.   As a result,  one should not need to use any regularization method at all to calculate the variation.  Shifts in divergent Feynman integrals, whose infinite parts cancel, must generate finite local expressions with derivatives.  This is illustrated in the literature, for example in \ci{taylor}.  The proof that this is so involves the proof that a theory is renormalizable (i.e. can be made finite, by changing the parameters in the action, by divergent amounts, depending on the regularization parameter), which is non-trivial, and we will not try to address that question here.  Assuming that this is so,  this can result in a conundrum.  If the local result cannot be obtained by a variation from a local action, then the result of the calulation is an anomaly of the theory. It cannot be removed by renormalization or by shifting divergent integrals, which is equivalent to finite renormalization.   This is the reason for the BRS cohomology analysis of  $\d_{\rm BRS}$, which looks at ghost charge one  cohomology (Form charge 5, if one  includes, as one  must, the integral as a Form charge 4 term  ) for local objects in the theory. This assumes that the variation can be characterized by a variation with ghost charge one , which is normally the case.  In particular the $\d_{\rm BRS}$ in  SUSY has this character.  

{\bf Antifields:} See Pseudofields.

{\bf BRS}: stands for Becchi-Rouet-Stora.  They realized that the Wess Zumino consistency condition   was an example of a general phenomenon, especially if one uses the fact that Fadeev Povov ghosts must be Grassmann odd.  Then they realized that this is really a cohomology problem, because the variation can be made to be nilpotent $\d_{\rm BRS}^2=0$ all cases, by using the structure functions of the theory for the ghosts.  The BRS operators which arise as the ``square roots' of the master equation in  section \ref{mastercss}  are  GO, since each term   has one  odd derivative and one  even derivative.  The BRS
 operator\footnote{Why is this called BRST by some authors?  While Tyutin certainly did some useful work, Tyutin had little or nothing to do with the ideas of BRS cohomology, as far as I know.}
for the present problem is  in equation (\ref{brstransWZ}). The literature is huge, and  \ci{poissonbrak,Becchi:1975nq,zinnbook,taylor,Zinnarticle,weinberg2,becchiarticles1, 
becchiarticles2} provides a start.

{\bf Chiral superfield:} This is the superspace version of the model of Wess and Zumino in \ci{WZ}.  We do not use superfields very often for the present work, particularly since the results are not superspace invariant \ci{dixexotic}. Superfields are very useful when they are relevant, but very deceptive when they are not.

{\bf Complex Conjugate:} This has the usual meaning for numbers and fields. We do not reverse the order of GO objects when taking the Complex Conjugate.  So no minus sign results from that.  Also we take the Action to be real, since our Feynman diagrams will be generated by path integration, not by canonical quantization.

{\bf Constraint equations:} These are described in \ci{dixexotic} and they appear frequently in this paper.

{\bf Construction of the \ED:} This starts out being rather easy for low dimensions and low spins.  It rapidly becomes increasingly difficult as the dimension and spin increase. In this paper we have stopped at dimension =1 with  $N_{\f} + N_{\ov \f}=1$ in section \ref{nightmare}.

{\bf CDSS:} \cdss: These are not treated in this paper.  They will be treated in a paper being prepared. They are the generalization of the constant spinor
$\f,\ov \f$ used here in section \ref{phiintro}.

{\bf $d_r$ 
Differential Operator}: See also $E_r$.  These were described extensively in \ci{dixspecseq} and they can be formed as described there.  In this article the denominators $\fr{1}{\D_0}$ that can sometimes appear in the definitions of $d_r,r\geq 2$ are simply numbers and they have been ignored.  It is conceivable that for higher dimensions, these might play a role. 
One very useful feature that frequently collapses the sequence, by ensuring that  $d_r=0,r\geq n$, so that $E_n = E_{\infty}$, is that each operator $d_r$ must satisfy the equation $\lt [ N_{\rm Grading}, d_r\rt ]= r d_r $.

{\bf Dimension}: The dimensions of various fields etc. are defined by the operator in section \ref{countingops}. 

{ \bf \ED:} We encounter this feature in the spectral sequence. See section  \ref{ghostquotes} for an explanation.   The book \ci{mccleary} has many examples of this, in a mathematical context. In this paper a number of these mappings can be found for example in section \ref{dimmin1nophi}.
In section \ref{quotesaboutspecseq} there are some quotes that summarize some of the features that are common when mathematicians  use spectral sequences to calculate topological quantities.

{\bf $E_r$ SubSpace}: These were described extensively in \ci{dixspecseq}. The initial nilpotent differential operator $\d_{\rm BRS}$ acts in the space $E_0$ of all local polynomials of the fields pseudofields and their derivatives and the ghosts.  The consecutive subspaces $E_{\infty}\subset \cdots E_{r+1} \subset E_r\cdots \subset E_0$ converge to a space $E_{\infty}$, which is isomorphic to the BRS cohomology of the operator  $\d_{\rm BRS}$ on the complete space $E_0$.  Usually one can expect  to be able to show that $d_r=0,r \geq n$ for some small positive integer n.  At each stage there is a new nilpotent differential $d_r$ acting on the space $E_r$, and  $d_r$ is formed by a procedure explained in  \ci{dixspecseq}.  Each successive space $E_{r+1}$ is the cohomology space of $d_r$.  See also the $d_r$ entry in this glossary.   

{\bf Exotic pair:} These can be found in section \ref{dimzerosec}.

{\bf Exotic triplet:} These can be found in section \ref{nightmare}.

 {\bf Fock Space}: See {\bf Adjoint} above. The idea here is that if we have  a local polynomial, say 
 \be
e_{a b} k^{(\a \g \d \dot \b)}  A^a_{\a \dot \b} \y^b_{\g} C_{\d}
\ee
where the tensors $e_{a b} k^{(\a \g \d \dot \b)}$ are dimensionless numerical tensors, 
we write it as follows 
 \be
\lt | V\rt >=
\lt | e_{a b} k^{(\a \g \d \dot \b)}A^a_{\a \dot \b} \y^b_{\g} C_{\d}\rt >
\ee
and its adjoint is
\be
\lt | V\rt >^{\dag} = \lt < V \rt |= \lt < (e_{a b} k^{(\a \g \d \dot \b)})^* A^{a\dag}_{\a \dot \b} \y^{b\dag} _{\g} C_{\d}^{\dag}\rt|
\ee
and clearly the inner product is positive semidefinite:
\be
\lt < V  |  V\rt > \geq 0
\ee
See \ci{dixspecseq} for further discussion.

 {\bf Form Charge}: This is ghost charge plus four.

{\bf Ghost Charge}: The ghost charge is defined by the operator  in section \ref{countingops}. Conventionally, ghost charge is equal to form charge minus four.  So anomalies have form charge five, and ghost charge one.  invariants have form charge four, and ghost charge zero. 

{\bf Grassmann Algebra}: This paper tries to use Greek letters for quantities that are Grassmann Odd ({ GO}) and Latin letters for quantities that are Grassmann even ({ GE}).
Thus for example, $C_{\a}, L_{\dot \a}, G_{\dot \a}, Y_{  \a}, A, F, W_{\a \dot \a},X_{\a \dot \a}$ are all GE and  their counterparts $\q_{\a}, \Lam, \G, \S_{\a \dot \a}, \y_{\a}, \f_{\dot \a},\c_{\dot \a}.\x_{\a\dot \a}$ are all GO. The BRS operators from the master equations in  section \ref{mastercss}    are  GO, since each term   has one  odd derivative and one even derivative.    

{\bf GE}: Grassmann Even

{\bf GO}: Grassmann Odd

{\bf GUT:} Grand Unified Theories \ci{ross,GUT}: These exotic pairs might be relevant to the supersymmetric GUTs.

{\bf Indices}: This paper uses both $i,j,k,\cdots $ and $a,b,c,\cdots $ for the indices on the various  fields and pseudofields, and the various constant dimensionless tensors that are contracted with them. This helps a bit to avoid using the same index improperly when pieces from different places are assembled together. The same happens with the spinor indices.

 {\bf Mass}:  This paper uses only  one  mass parameter to make the 
dimension consistent.  All the tensors here are dimensionless. 

{\bf Master equation}: The BRS operators from the master equation in  section \ref{mastercss}  are  GO, since each term   has one odd derivative and one even derivative.    

{\bf Missing Terms in the Elizabethan drama}: This is a concern, certainly.  One needs to check this work repeatedly. This is one of the features that makes the \ED\ obscure and difficult.  It is somewhat of a surprise how well it works.  This is buried in the amazingly complicated definition of the spectral sequence. 

 {\bf Pseudofields:} This term  is used here for  the sources for the variations of the fields. These were originally introduced   by J Zinn Justin. These Zinn-Justin sources are not fields, and they are certainly not antifields, either. Moreover, the antifields of the fields (their complex conjugates) play an important role in the cohomology in this paper.  The Zinn-Justin sources do not get quantized. So in this paper, we use the name pseudofields to refer to these Zinn-Justin sources.  They were originally introduced by Zinn-Justin to formulate his identity, which was later named the `master equation'. Later they were renamed `antifields' by a number of authors. Here we call them \Pf s because they are sources, and certainly not fields, and they are certainly not identical to the complex conjugates of fields. The name  `master equation' seems useful however, although really it is a kind of Poisson Bracket, which is invariant under canonical transformations, as modified by the Grassmann nature of the fields and pseudofields.

{\bf Signs, Factors, Conventions and Errors}: These can certainly be an  irritating, time consuming, and disturbing problem in  supersymmetry.  For example, the authors of { Superspace}  \ci{superspace} state that ``if the reader thinks he sees an error, he is probably right''.  Note that the spectral sequence is very  forgiving in this regard.  So long as there really is a nilpotent $\d_{\rm BRS}$ with roughly (up to signs and factors) the form in equations (\ref{brstransWZ}), the spectral sequence is very insensitive to the correctness of the signs and factors.   While I have tried to be accurate in regard to signs and factors, it usually does not matter much for the present problem.

{\bf  \Sps:} This is introduced for BRS in \ci{dixspecseq}.

{\bf Structure Functions for the ghosts}:  For a non-Abelian gauge theory we have the expression $f_{abc} \w^a\w^b\w^c$ and for SUSY we have 
$C^{\a} \oC^{\dot \b}\x_{\a \dot \b} $.  The derived nilpotent ghost operators for these two are  $\d_{\rm Gauge\;ghosts}= f_{abc} \w^a\w^b \fr{\pa}{\pa \w^c}$ and $\d_{\rm SUSY\;ghosts}=C^{\a} \oC^{\dot \b}\fr{\pa}{\x_{\a \dot \b} }$. The cohomology of $\d_{\rm SUSY\;ghosts}$ is crucial in this paper.  It is the basis of the formula (\ref{structureofE1line3}) for $E_1$.
The (much simpler) cohomology of $\d_{\rm Gauge\;ghosts}$ is also crucial for gauge theories.

{\bf Superstring}: Once the superstring \ci{witten} is reduced to 3 + 1 dimensions, it may be that the exotic objects here are of interest, including the ones with higher spin.

{\bf SUSY}: This is an acronym for supersymmetry. 
Some introductions are contained in \ci{WZ,weinbergcosmo,west,freepro,ferrarabook,superspace,Buchbinder:1998qv,xerxes,Weinberg3,haber,buchmueller,bagger,witten,pran}.

{\bf WZ:} stands for Wess and Zumino, and their chiral supersymmetry model which is described in \ci{WZ}. Later it was discovered to be a model of the chiral superfield. 

{\bf Zinn-Justin Sources and Identity}: Many of his important contributions can be found in \ci{Zinnarticle}.   See \ci{dixonnucphys} for an early reference with a focus on renormalization and the analogy to the Poisson Bracket and canonical transformations.

\begin{center}
 { Acknowledgments}
\end{center}
\vspace{.1cm}

  I thank  Doug Baxter,  Carlo Becchi,  Margaret Blair, Friedemann Brandt, Philip Candelas, David Cornwell,   James Dodd, Mike Duff, Sergio Ferrara, Richard Golding, Dylan Harries, Marc Henneaux, Chris T.  Hill,  D.R.T. Jones, Antoine van Proeyen,  Pierre Ramond,   Peter Scharbach,      Mahdi Shamsei, Kelly Stelle, Sean Stotyn, Xerxes Tata, J.C. Taylor,  Peter West and Ed Witten for stimulating correspondence and conversations.   I also express appreciation for help in the past from Lochlainn O'Raifeartaigh, Graham Ross, Raymond Stora and Steven Weinberg. They are not replaceable and they are missed.  I particularly thank Margaret Blair, David Cornwell, James Dodd, Peter Scharbach and  Pierre Ramond for recent, and helpful, encouragement to carry on with this work.

\begin{center}
 {Declarations}
\end{center}
\vspace{.1cm}

 Funding and Competing Interests: The author did not receive support from any organization for this work. The author has no competing interests to declare that are relevant to the content of this article.

Data availability statement:
 There is no data associated with this paper.

 \tableofcontents

\tiny 
\articlenumber\\\today

\end{document}

%% file: Johnmacros27.tex
\newcommand{\Pf}{{pseudofield}} 
\newcommand{\oCt}{{\rm (\oC \x^2 \oC)  }} 
\newcommand{\Ct}{{\rm (C \x^2 C)  }}

  \newcommand{\cdss}{chiral dotted spinor superfield}
  
 \newcommand{\ED}{Elizabethan drama}
\newcommand{\wzcss}{Wess Zumino Chiral Scalar Superfield}

\newcommand{\cC}{{\cal C}}
\newcommand{\cE}{{\cal E}}

\newcommand{\cA}{{\cal A}}

\newcommand{\CC}{Complex Conjugate}

\newcommand{\Exists}{\bm{\exists}\kern-0.6em\bm{\exists}}

\newcommand{\PB}{Master Equation}

\newcommand{\SSM}{Supersymmetric Standard Model}

\newcommand{\Sps}{Spectral Sequence}

\newcommand{\bt}{\begin{tabular}{c}}
\newcommand{\et}{\end{tabular}}

\newcommand{\eb}{\ee\be } 
\newcommand{\ebp}{\rt.\ee\be\lt.} 
\newcommand{\bmat}{\lt ( \begin{array} }
\newcommand{\emat}{  \end{array} \rt )}

\newcommand{\oP}{{\ov P}}

\newcommand{\oQ}{{\ov Q}}

\newcommand{\oR}{{\ov R}}

\newcommand{\cH}{{\cal H}}
\newcommand{\cP}{{\cal P}}
\newcommand{\cF}{{\cal F}}

\newcommand{\oq}{{\ov \q}}
\newcommand{\oT}{{\ov T}}
\newcommand{\oG}{{\ov \G}}

\newcommand{\oY}{{\ov Y}}
\newcommand{\og}{{\ov g}}
\newcommand{\oy}{{\ov \y}}

\newcommand{\om}{{\ov m}}

\newcommand{\oC}{{\ov C}}

\newcommand{\oF}{{\ov F}}
\newcommand{\A}{{\ov A}}

\renewcommand{\a}{\alpha}	
\renewcommand{\b}{\beta}
\newcommand{\g}{\gamma}
\renewcommand{\d}{\delta}
\newcommand{\e}{\epsilon}
\newcommand{\ve}{\varepsilon}
\newcommand{\z}{\zeta}
\newcommand{\h}{\eta}
\newcommand{\q}{\theta}

\newcommand{\lam}{\lambda}
\newcommand{\m}{\mu}
\newcommand{\n}{\nu}	
\newcommand{\x}{\xi}

\newcommand{\s}{\sigma}

\newcommand{\f}{\phi}
\renewcommand{\c}{\chi}
\newcommand{\y}{\psi}
\newcommand{\w}{\omega}
\newcommand{\G}{\Gamma}
\newcommand{\D}{\Delta}
	
\newcommand{\Lam}{\Lambda}
\newcommand{\X}{\Xi}
\renewcommand{\P}{\Pi}	
\renewcommand{\S}{\Sigma}

\newcommand{\W}{\Omega}
\newcommand{\la}{\label}
\newcommand{\ci}{\cite}

\newcommand{\ds}{\documentstyle}	
\newcommand{\fr}{\frac}
\newcommand{\na}{\nabla}
\newcommand{\pa}{\partial}
\newcommand{\ov}{\overline}
\newcommand{\br}{\begin{rant}}
\newcommand{\er}{\end{rant}}
\newcommand{\beC}{\begin{Conjecture}}
\newcommand{\eeC}{\end{Conjecture}}
\newcommand{\be}{\begin{equation}}
\newcommand{\ee}{\end{equation}}
\newcommand{\ba}{\begin{array}} 
\newcommand{\ea}{\end{array}}
\newcommand{\bea}{\begin{eqnarray}}
\newcommand{\eea}{\end{eqnarray}}
\newcommand{\ra}{\rightarrow}
\newcommand{\Ra}{\Rightarrow}
\newcommand{\lra}{\longrightarrow}
\newcommand{\lla}{\longleftarrow}

\newcommand{\lt}{\left}
\newcommand{\rt}{\right}

\newcommand{\ben}{\begin{enumerate}}
\newcommand{\een}{\end{enumerate}}
\newcommand{\bitem}{\begin{itemize}}
\newcommand{\eitem}{\end{itemize}}